\documentclass[letter,11pt]{article}
\def\fighome {figures/}

\usepackage{fullpage}

\usepackage{amsmath}
\usepackage{epsfig,epsf,psfrag,amssymb,amsfonts,latexsym,slashbox,graphicx,bm,cite,xcolor,url,array}
\usepackage[caption=false]{subfig}
\usepackage{fixltx2e}
 \usepackage{cases}
\usepackage{verbatim}
\usepackage[mathscr]{eucal}
\usepackage{algorithm, algorithmic}
\usepackage[fleqn,tbtags]{mathtools}

   \DeclareMathOperator{\Path}{Path}

 \DeclareMathOperator{\Deg}{Deg}

\DeclareMathOperator{\Diam}{Diam}

 \def\0{{\bf 0}}

\def\nn{\nonumber}

\def\qed{\hfill\hbox{${\vcenter{\vbox{
    \hrule height 0.4pt\hbox{\vrule width 0.4pt height 6pt
    \kern5pt\vrule width 0.4pt}\hrule height 0.4pt}}}$}}

\definecolor{myred}{rgb}{0.3,0.0,0.7}
\definecolor{dkg}{rgb}{0.1,0.7,0.2}
\definecolor{dkb}{rgb}{0.0,0.2,0.8}

\newcommand{\Gmsc}{\mathscr{G}}

 \def\hl{\widehat{l}}

 \def\hG{\widehat{G}}

\def\bfA{{\mathbf A}}

\def\bfD{{\mathbf D}}

\def\Bc{{\cal B}}
\def\Cc{{\cal C}}
\def\Dc{{\cal D}}
\def\Ec{{\cal E}}

\def\Gc{{\cal G}}

\def\Qc{{\cal Q}}
\def\Rc{{\cal R}}
\def\Sc{{\cal S}}

\def\Wc{{\cal W}}

\def\Ebb{{\mathbb E}}

\def\Ibb{{\mathbb I}}

\def\Nbb{{\mathbb N}}

\def\Pbb{{\mathbb P}}

\newcommand{\bprf}{\begin{myproof}}
\newcommand{\eprf}{\end{myproof}}
\newcommand{\bp}{\begin{psfrags}}
\newcommand{\ep}{\end{psfrags}}
\newcommand{\bl}{\begin{lemma}}
\newcommand{\el}{\end{lemma}}
\newcommand{\bt}{\begin{theorem}}
\newcommand{\et}{\end{theorem}}
\newcommand{\bc}{\begin{center}}
\newcommand{\ec}{\end{center}}
\newcommand{\bi}{\begin{itemize}}
\newcommand{\ei}{\end{itemize}}
\newcommand{\ben}{\begin{enumerate}}
\newcommand{\een}{\end{enumerate}}
\newcommand{\bd}{\begin{definition}}
\newcommand{\ed}{\end{definition}}
\def\beq{\begin{equation}}
\def\eeq{\end{equation}\noindent}
\def\beqn{\begin{eqnarray}}
\def\eeqn{\end{eqnarray} \noindent}
\def\beqnn{  \begin{eqnarray*}}
\def\eeqnn{\end{eqnarray*}  \noindent}
\def\bcase{  \begin{numcases}}
\def\ecase{\end{numcases}   \noindent}
\def\bsbcase{  \begin{subnumcases}}
\def\esbcase{\end{subnumcases}   \noindent}

\newtheorem{theorem}{Theorem}
\newtheorem{corollary}{Corollary}
\newtheorem{lemma}{Lemma}

\newtheorem{definition}{Definition}

\newenvironment{myproof}{\noindent{\em Proof:} \hspace*{1em}}{
    \hspace*{\fill} $\Box$ }
\newenvironment{proof_of}[1]{\noindent {\em Proof of #1: }}{\hspace*{\fill} $\Box$ }

\newcommand{\matplottc}[1]{               
        \unitlength .45truein
        \begin{center}
        \includegraphics{#1.ps}
        \end{picture}
        \end{center}
}

\def\psfancypar#1#2{\begingroup\def\par{\endgraf\endgroup\lineskiplimit=0pt}
               \setbox2=\hbox{\large\sc #2}
               \newdimen\tmpht \tmpht \ht2 \advance\tmpht by \baselineskip
               \font\hhuge=Times-Bold at \tmpht
               \setbox1=\hbox{{\hhuge #1}}
               \count7=\tmpht \count8=\ht1
               \divide\count8 by 1000 \divide\count7 by \count8
               \tmpht=.001\tmpht\multiply\tmpht by \count7
               \font\hhuge=Times-Bold at \tmpht
               \setbox1=\hbox{{\hhuge #1}}
               \noindent
                \hangindent1.05\wd1
               \hangafter=-2 {\hskip-\hangindent
               \lower1\ht1\hbox{\raise1.0\ht2\copy1}%
                \kern-0\wd1}\copy2\lineskiplimit=-1000pt}

\def\Kout{\setbox1=\hbox{\Huge\bf K}\hbox to
1.05\wd1{\hspace{.05\wd1}
}}
\def\Sout{\setbox1=\hbox{\Huge\bf S}\hbox to 1.05\wd1{\hspace{.05\wd1}
}}

\newcommand{\tilG}{\widetilde{G}}

\newcommand{\tilbD}{\widetilde{\mathbf{D}}}

\allowdisplaybreaks[4]

\newcommand{\bD}{\mathbf{D}}

\newcommand{\bN}{\mathbb{N}}

\newcommand{\poly}{\mathrm{poly}}

\newcommand{\estone}{\mathsf{RGD1}}
\newcommand{\esttwo}{\mathsf{RGD2}}
\newcommand{\mergequartet}{\mathsf{QuartetMerge}}
\newcommand{\TreeMerge}{\mathsf{TreeMerge}}
\newcommand{\CycleMerge}{\mathsf{CycleMerge}}
\newcommand{\minimal}{\mathsf{Minimal}}
\DeclareMathOperator{\bad}{bad}

\begin{document}

\title{Topology Discovery of Sparse Random Graphs
 \\With Few Participants\footnote{A shorter version appears in \cite{Anandkumar&etal:Sigmetrics11}. This version is scheduled to appear in Journal on Random Structures and Algorithms.}}%

\author{Animashree Anandkumar\footnote{A. Anandkumar is with the Center for Pervasive Communications and Computing, Electrical Engineering and Computer Science Dept., University of California, Irvine, USA 92697. Email: a.anandkumar@uci.edu}, Avinatan Hassidim\footnote{
A. Hassidim is with Google Research,  Tel Aviv, Israel. Email: avinatanh@gmail.com} and Jonathan Kelner\footnote{J. Kelner is with the Computer Science and Artificial Intelligence Laboratory, Massachusetts Institute of Technology, Cambridge MA 02139.  Email: kelner@mit.edu}}

\maketitle

\begin{abstract}We consider the task of topology discovery of  sparse random graphs using end-to-end random measurements (e.g., delay) between a subset of nodes, referred to as the participants. The rest of the nodes are hidden, and do not provide any information for topology discovery.  We consider topology discovery under two routing models: (a) the participants exchange messages along the shortest paths  and obtain end-to-end measurements, and (b) additionally, the participants exchange messages along the second shortest path.  For scenario (a), our proposed algorithm results in a sub-linear edit-distance guarantee using  a sub-linear number of uniformly selected participants. For scenario (b), we obtain a much stronger result, and show that we can achieve consistent reconstruction when a  sub-linear number of uniformly selected nodes participate. This implies that  accurate  discovery of sparse random graphs is tractable using an extremely small number of participants.   We finally obtain a lower bound on the  number of participants  required by any algorithm to reconstruct the original random graph up to a given edit distance. We also demonstrate that while consistent  discovery is tractable for sparse random graphs using a small number of participants, in general, there are graphs which cannot be discovered by any algorithm even with a significant number of participants, and with the availability of end-to-end information along all the paths  between the participants.
\end{abstract}






\section{Introduction}

Inference of global characteristics of large networks using limited local information is an important and a challenging task. The discovery of the underlying network  topology  is one of the main goals of network inference, and its knowledge  is crucial for many  applications. For instance, in communication networks,  many network monitoring applications rely on the knowledge of the routing topology, e.g., to evaluate  the resilience of the network to failures~\cite{Kandula&etal:05Sigcomm,Motter&Lai:02APS}; for network traffic prediction~\cite{Eriksson&etal:07Sigcomm,Vardi:96ASA} and monitoring~\cite{Anandkumar&Bisdikian&Agrawal:08Sigmetrics}, anomaly detection~\cite{Alderson&etal:06AIMS},   or to infer the sources of viruses and rumors in the network~\cite{Shah&Zaman:10Sigmetrics}. In the context of social networks, the knowledge of topology  is useful for inferring many characteristics such as identification of hierarchy and community structure \cite{Fortunato:09Phy}, prediction of information flow \cite{Wu&etal:04Phy,Acemoglu&etal:10Games}, or to evaluate the possibility of information leakage from  anonymized social networks~\cite{Backstrom:07WWW}.

Traditionally, inference of  routing topology in communication networks has relied on tools such as traceroute and mtrace~\cite{mtrace} to generate path information between a subset of nodes. However, these tools require cooperation of intermediate nodes or routers to  generate messages using the Internal Control Message Protocol (ICMP). Increasingly, today many routers  block traceroute requests due to privacy and security concerns \cite{Gunes&etal:08INFOCOM,Yao&etal:03INFOCOM}, there by making inference of topology using traceroute inaccurate. Moreover,  traceroute requests are not scalable for large networks, and cannot discover   layer-2 switches and MPLS (Multi-protocol Label Switching) paths, which are increasingly being deployed~\cite{Ni&etal:10TON}.

The alternative approach for topology discovery  is the approach of {\em network tomography}. Here, topology inference is carried out  from end-to-end packet probing measurements (e.g., delay)  between a subset of nodes, without the need for cooperation between the intermediate (i.e., non-participating) nodes in the network. Due to its flexibility, such approaches are gaining increasing popularity (see Section~\ref{sec:related} for details).

The approach of topology discovery using end-to-end measurements  is also applicable in the context of social networks. In many social networks, some nodes may be unwilling to participate or cooperate with other nodes for discovering the network topology, and there may be many hidden nodes in ``hard to reach'' places of the network, e.g., populations of drug users, and so on. Moreover, in many networks, there may be a cost to probing  nodes for information, e.g., when there is a cash reward offered for filling out surveys. For such networks, it is desirable to design algorithms which can discover the overall network topology using small fraction of participants who are willing to provide information for topology discovery.

There are many challenges to topology discovery. The algorithms need to be computationally efficient and provide accurate reconstruction using a small fraction of participating nodes. Moreover, inference of large topologies is a task of {\em high-dimensional learning}~\cite{Wainwright&Jordan:08NOW}. In such scenarios, typically, only a small number of end-to-end measurements are available relative to the size of the network to be inferred. It is desirable to have algorithms with low {\em sample complexity} (see Definition~\ref{def:samplecomplexity}), where the number of measurements required to achieve a certain level of accuracy scales favorably with the network size.

It is  indeed not  tractable to achieve all the above objectives for discovery of   general network topologies using an arbitrary set of participants. There are fundamental identifiability issues, and in general, no algorithm will be able to discover the underlying topology. We demonstrate this phenomenon in Section~\ref{sec:nonidentifiable}, where we construct a small network with a significant fraction of participants which suffers from non-identifiability. Instead, it is desirable to design topology discovery algorithms which have guaranteed performance for certain classes  of graphs.

We consider the class of Erd\H{o}s-R\'{e}nyi random graphs \cite{Bollobas:book}. These are perhaps the simplest as well as the most well-studied class of random graphs. Such random graphs can provide a reasonable  explanation for peer-to-peer networks~\cite{Jovanovic:01} and social networks~\cite{Newman:02}. We address the following issues in this paper: can we discover random graphs using a small fraction of participating nodes, selected uniformly at random? can we design efficient algorithms with low sample complexity and with provable performance guarantees? what kinds of end-to-end measurements between the participants are useful for topology discovery? finally, given a set of participants,  is there a lower bound on the error (edit distance) of topology discovery   that is achievable by any algorithm? Our work addresses these questions and also provides insights into many complex issues involved in topology discovery.

\subsection{Summary of Contributions}

We consider the problem of topology discovery of sparse random graphs using a uniformly selected set of participants. Our contributions in this paper are three fold. First, we design an algorithm  with provable performance guarantees, when only minimal end-to-end information  between the participants is available. Second, we consider the scenario with additional information, and design a discovery algorithm with much better reconstruction guarantees. Third, we provide a lower bound on the edit distance of the reconstructed graph by any algorithm, for a given number of participants.
 Our analysis shows that random graphs can be discovered accurately and efficiently using an extremely small number of participants.

We consider reconstruction of the giant component of the  sparse random graph  up to its minimal representation, where there are no redundant hidden    nodes  (see Section~\ref{sec:minimal}).
Our end-to-end measurement model consists of random samples (e.g., delay) along the shortest paths between the participants. Using these   samples, we design the first random-discovery algorithm, referred to as the $\estone$ algorithm, which performs local tests over small groups of participating nodes (known as the {\em quartet tests}), and iteratively merges them with the previously constructed structure. Such tests are known to be accurate for tree topologies~\cite{Bhamidi&etal:09Rand}, but have not been previously analyzed for random-graph topologies. We provide a sub-linear edit-distance guarantee (in the number of nodes) under $\estone$ when there are roughly   $n^{5/6}$   participants,  where $n$ is the number of nodes in the network. The algorithm is also simple to implement, and is computationally efficient.

We then extend the algorithm   to the scenario where additionally, there are end-to-end measurements available along the second shortest paths between the participating nodes. Such information is available since nodes typically maintain information about alternative routing paths, should the shortest path fail. In this scenario, our algorithm $\esttwo$, has a drastic improvement in accuracy under the same set of participating nodes. Specifically, we demonstrate that consistent discovery  can be  achieved under $\esttwo$ algorithm when there are roughly $n^{11/12}$    number of participants, where $n$ is the network size. Thus, we can achieve accurate topology discovery of random graphs using an extremely small number of participants.
For both our algorithms, the sample complexity is poly-logarithmic in the network size, meaning that the number of end-to-end measurement samples needs to scale poly-logarithmically in the network size to obtain the stated edit-distance guarantees.

Our analysis in this paper thus reveals that sparse random graphs can be efficiently discovered using a small number of participants. Our algorithms exploit the {\em locally tree-like} property of random graphs~\cite{Bollobas:book}, meaning that these graphs contain a small number of short cycles. This  enables us to provide performance guarantees for   quartet tests which are known to be accurate for tree topologies, and this is done by carefully controlling the distances used by the quartet tests. At the same time, we exploit the presence of cycles in random graphs to obtain much better guarantees than in the case of tree topologies. In other words, while tree topologies require participation of at least half the number of nodes (i.e., the leaves) for accurate discovery, random-graph topologies can be accurately discovered using a sub-linear number of participants.




Finally, we provide lower bounds on the reconstruction error under any algorithm for a given number of participants. Specifically, we show that if less than roughly $\sqrt{n}$ nodes participate in topology discovery, reconstruction is impossible under any algorithm, where $n$ is the network size. We also discuss topology discovery in general networks, and demonstrate identifiability issues involved in the discovery process. We construct a small network with a significant fraction of nodes as participants which cannot be reconstructed using   end-to-end information on all possible paths between the participants. This is in contrast to random graphs, where consistent and efficient topology discovery is possible  using a small number of participants.

To the best of our knowledge, this is the first work to undertake a systematic study of random-graph discovery using end-to-end measurements between a subset of nodes.
Although we limit ourselves to the study of random graphs, our algorithms are based on the locally tree-like property, and are thus equally  applicable for discovering other locally tree-like graphs such as the $d$-regular graphs and the {\em scale-free} graphs; the latter class is   known to be a good model for social networks~\cite{barabási1999emergence,Newman:02} and peer-to-peer networks~\cite{Jovanovic:01}. Indeed more sophisticated  and general models for networks have been developed~\cite{leskovec2010kronecker,leskovec2008statistical,clauset2008hierarchical}, but we defer their study for future work.


\subsection{Related Work}\label{sec:related}

Network tomography has been extensively studied in the past and various heuristics and algorithms have been proposed along with experimental results on real data. For instance, the area of mapping the internet topology is very rich and extensive, e.g., see \cite{internetmap,skitter,caida,Govindan:00Infocom,Eriksson&etal:07Sigcomm,Spring&etal:04TON,Shavitt:05Sigcomm,He&etal:bookchapter}. In the context of social networks, the work in~\cite{Leskovec&etal:10WWW} considers  prediction of positive and negative links, the work in~\cite{Gomez&etal:10KDD} considers inferring networks of diffusion and influence and the work in~\cite{Myers&Leskovec:10NIPS} considers inferring latent social networks through spread of contagions. A wide range of network tomography solutions have been proposed for general networks. See~\cite{Castro04} for a survey.

Topology discovery is an important component of network tomography. There have been several theoretical developments on this topic. The work in~\cite{Chung&etal:CSS} provides hardness results for topology discovery under various settings. Topology discovery under availability of different kinds of queries  have been previously   considered, such as:

\noindent{\em (i) Shortest-path query}, where a query to a node returns all the shortest paths (i.e., list of nodes in the path) from that node to all other nodes~\cite{Beerliova&etal:06JSAC}. This is the strongest of all queries.  These queries can be implemented by using Traceroute on Internet. In \cite{Beerliova&etal:06JSAC}, the combinatorial-optimization problem of selecting the smallest subset of nodes for such queries to estimate the network topology is formulated. The work in \cite{Erlebach&etal:07LNCS} considers discovery of random graphs using such queries. The bias of using traceroute sampling on power-law graphs is studied in~\cite{AchlioptasCKM09}, and weighted random walk sampling is considered in~\cite{Kurant:Sigmetrics11}.

\noindent{\em (ii) Distance query}, where a query to a node returns all the shortest-path distances (instead of the complete list of nodes) from that node to any other node in the network~\cite{Erlebach&etal:07LNCS}. These  queries are available   for instance, in Peer-to-Peer networks through the Ping/Pong protocol. This problem is   related to the landmark placement, and the optimization problem of having smallest number of landmarks is known as the metric dimension of the graph \cite{Khuller&etal:DM96}. The work in~\cite{Reyzin&Srivastava:07IPL} considers reconstruction of tree topologies using shortest-path queries.

\noindent{\em (iii) Edge-based queries:} There are several types of edge queries such as detection query, which answer whether there is an edge between two selected nodes, or counting query, which returns number of edges in a selected subgraph~\cite{Reyzin&Srivastava:07LNCSS,Mazzawi:10SODA}, or a  cross-additive query, which returns the number of edges crossing between two disjoint sets of vertices~\cite{Choi&Kim:08STOC}.

However, all the above queries assume that all the nodes (with labels) are known a priori, and  that there are no hidden (unlabeled) nodes in the network. Moreover, most of the above works consider unweighted graphs, which are not suitable when end-to-end delay (or other weighted) information is available for topology discovery. As previously discussed, the above queries assume extensive information is available from the queried objects, and this may not be feasible in many networks.

Topology discovery using end-to-end delays between a subset of nodes (henceforth, referred to as participating nodes), has been previously studied for tree topologies using unicast traffic in~\cite{Bhamidi&etal:09Rand,Ni&etal:10TON,Shih:02ICASSP} and multicast traffic~\cite{Duffield&etal:00PA}. The algorithms are  inspired by {\em phylogenetic} tree algorithms. See~\cite{Durbin} for a thorough review.
Most of these algorithms are based on a series of local tests known as the    {\em quartet-based  distance tests}. Our algorithms are inspired by, and are based on quartet methods. However,   these algorithms
were previously applied only  to tree topologies, and here, we show how algorithms based on similar ideas can provide accurate reconstruction  for a much broader class of locally-tree like graphs such as the sparse random graphs. Recent works also incorporate additional information from temporal dynamics~\cite{gomez2011uncovering} or consider causal  models for networks~\cite{lappas2010finding,snowsill2011refining}, while our work does not consider these effects.



\section{System Model}

\subsubsection*{Notation}

For any two functions $f(n),g(n)$, $f(n) = O(g(n))$ if there exists a constant $M$ such that $f(n) \leq M g(n)$ for all $n \geq n_0$ for some fixed $n_0\in \Nbb$. Similarly, $f(n) = \Omega(g(n))$ if there exists a constant $M'$ such that $f(n) \geq M'g(n)$  for all $n \geq n_0$ for some fixed $n_0\in \Nbb$, and $f(n) = \Theta(g(n))$ if $f(n)= \Omega(g(n))$ and $f(n) = O(g(n))$. Also, $f(n) = o(g(n))$ when $f(n)/g(n) \to 0$ and $f(n) = \omega(g(n))$ when $f(n)/ g(n) \to \infty$ as $n \to \infty$. We use notation $\tilde{O}(g(n)) = O(g(n)\poly \log n)$. Let $\Ibb[A]$ denote indicator of an event $A$.

Let  $G_n $ denote a  random graph with probability measure $\Pbb$.  Let $\Qc$ be a  graph property (such as being connected). We say that the property $\Qc$ for a sequence of random graphs $\{G_n\}_{n\in\bN}$ holds asymptotically almost surely (a.a.s.) if, \[\lim_{n \to \infty}\Pbb(G_n\,\, \mathrm{satisfies} \,\, \Qc)= 1.\]Equivalently,  the property $\Qc$ holds for {\em almost every} (a.e.) graph $G_n$.


For a graph $G$, let $\Cc(l;G)$ denote the set of (generalized) cycles\footnote{A generalized cycle of length $l$ is a connected graph of $l$ nodes with $l$ edges (i.e., can be a union of a path and a cycle). In this paper, a cycle refers to a generalized cycle unless otherwise mentioned.} of length less than $l$ in graph $G$. For a vertex $v$, let $\Deg(v)$ denote its degree and for an edge $e$, let $\Deg(e)$ denote the total number of edges connected to either of its endpoints (but not counting the edge $e$). Let $ B_R(v)$ denote the set of nodes within hop distance $R$ from a node $v$ and $\Gamma_R(v)$ is the set of nodes exactly at hop distance $R$. The definition is extended to an edge, by considering union of sets of the endpoints of edge. Denote the shortest path (with least number of hops) between two nodes $i, j$ as $\Path(i,j;G)$ and the second shortest path as $\Path_2(i,j;G)$. Denote the number of $H$-subgraphs in $G$, i.e., the number of subgraphs in $G$ corresponding to $H$,  as $N_{H;G}$.

\subsection{Random Graphs}

We assume that the unknown network topology is drawn from the ensemble of Erd\H{o}s-R\'{e}nyi random graphs \cite{Bollobas:book}.  This random graph model  is arguably the simplest as well  the most well-studied model. Denote the random graph as $G_n \in \Gmsc(n,c/n)$, for $c<\infty$, where  $n$ is the number of nodes and each edge occurs uniformly with probability $c/n$. This implies a constant average degree of $c$ for each node, and   this regime is also known as the ``sparse'' regime of random graphs.

It is well known that sparse random graphs exhibit a phase transition with respect to the number of components. When $c>1$, there is a giant component   containing $\Theta(n)$ nodes, while all the other components have size $\Theta(\log n)$ \cite[Ch. 11]{Alon:book}. This regime is known as the {\em super-critical} regime.  On the other hand, when $c<1$, there is no giant component and all components have size $\Theta(\log n)$. This regime is known as the {\em sub-critical} regime.

We consider discovery of a random graph in the  super-critical regime $(c>1)$. This is the regime of interest, since most real-world networks are well connected rather than having large number of extremely small components. Moreover, the presence of a giant component   ensures that the topology can be discovered even with a small fraction of random participants. This is because the participants will most likely belong the giant component, and can thus exchange messages between each other to discover the unknown topology. We limit ourselves to the topology discovery of the giant component in the random graph, and denote the giant component as $G_n$, unless otherwise mentioned.


\subsection{Participation Model}

For the given unknown graph topology  $G_n=(W_n, E_n)$ over $W_n=\{1,\ldots, n\}$ nodes, let $V_n\subset W_n$ be the set of participating   nodes which exchange messages amongst each other by routing them along the graph. Let $\rho_n :=\frac{|V_n|}{n}$ denote the fraction of participating nodes. It is desirable to have small $\rho_n$ and still reconstruct the unknown topology.
We assume that the nodes decide to participate uniformly at random.  This ensures that  information about all parts of the graph can be obtained, thereby  making graph reconstruction feasible. We consider the regime, where $  |V_n| = n^{1 - \epsilon}$, for some $\epsilon > 0$, meaning that extremely small number of nodes participate in discovering the topology.

Let $H_n :=W_n \setminus V_n$ be the set of hidden nodes. The hidden nodes only forward the messages without altering them, and   do not provide any additional information for topology discovery. The presence of hidden nodes thus needs to be inferred, as part of our goal of discovering the unknown graph  topology.

\subsection{Delay Model}\label{sec:delaymodel}

The messages  exchanged between the participating nodes experience  delays along the links in the route. The participating nodes measure the end-to-end delays\footnote{Our algorithms work under any {\em additive metric} defined on the graph such as    link utilization or link loss~\cite{Ni&etal:10TON}, although the sample complexity, i.e., the number of samples required to accurately estimate the metrics, does indeed depend on the metric under consideration.} between message transmissions and receptions. We consider the challenging scenario that  only this end-to-end delay information is available for topology discovery.

Let $m$ be the number of messages exchanged between each pair of participating nodes $i,j \in V_n$.
Denote the $m$ samples of end-to-end delays computed from these messages   as \[\bD^m_{i,j}:=[D_{i,j}(1), D_{i,j}(2),\ldots, D_{i,j}(m)]^T.\] We assume that the routes taken by the $m$ messages are fixed, and we discuss the routing model in the subsequent section. On the other hand, these messages experience different  delays along each link\footnote{The independence assumption implies that we consider unicast traffic rather than multicast traffic considered in many other works,~e.g., in~\cite{Duffield&etal:00PA}.} which are drawn identically and independently  (i.i.d) from some distribution, described below.

Let $D_e$ denote the random  delay along a link $e\in G_n$ (in either direction). We assume that the delays $D_{e_1}$ and $D_{e_2}$ along any two links $e_1,e_2\in G_n$ are independent. The  delays are additive along any  route, i.e., the end-to-end delay along a route $\Rc(i,j)$ between two participants $i,j\in V_n$ is\beq D_{\Rc(i,j)}:= \sum_{e\in \Rc(i,j)} D_e.\label{eqn:additive}\eeq
Further, the family of delay distributions are regular and bounded, as in \cite{Bhamidi&etal:09Rand}.

The delay distributions $\{D_e\}_{e\in E_n}$ and the graph topology $G_n$ are both unknown, and need to be estimated using messages between participating nodes. We   exploit the additivity assumption in \eqref{eqn:additive} to obtain efficient topology discovery algorithms.

\subsection{Routing Model}\label{sec:scenarios}

The end-to-end delays  between the participating nodes thus depends on the routes taken by the messages. We assume that the messages between any two participants are routed along the shortest path with the lowest number of hops. On the other hand, the nodes cannot select the path with the least delay since the delays along the individual links are unknown and are also different for different messages.

We also consider another scenario, where the participants are able to additionally route messages along the second shortest path. This  is   a reasonable assumption, since  in practice, nodes typically maintain  information about the shortest path and an alternative path, should the shortest path fail. The nodes can forward messages along the shortest and the second shortest paths with different headers, so that the destinations can distinguish the two messages and compute the end-to-end delays along the two paths. We will show that this additional information vastly improves the accuracy of topology discovery. These two scenarios are formally defined below.

\noindent{\bf Scenario 1 (Shortest Path Delays): }Each pair of participating nodes $i,j \in V_n$ exchange $m$   messages along the shortest path  in $G_n$, where the shortest path\footnote{If the shortest path between two nodes is not unique, assume that the node pairs randomly pick one of the paths and use it for all the messages.} is with respect to the number of hops. Denote the vector of $m$ end-to-end delays as $\bD^m_{i,j}$.

\noindent{\bf Scenario 2 (Shortest Path and Second Shortest Path Delays): }Each pair of participating nodes $i,j \in V_n$ exchange $m$  messages along the shortest path   as well as $m$   messages along the second shortest path.  The vector of $m$ samples along the second shortest path is denoted by $\tilbD^m_{i,j}$.

\section{Reconstruction Guarantees}

\subsection{Minimal Representation}\label{sec:minimal}

Our goal is to discover the unknown graph topology using the end-to-end delay information between the participating nodes. However, there can be multiple topologies which  explain equally well the  end-to-end delays  between the participants. This inherent ambiguity in topology discovery with hidden nodes has been previously pointed out in the context of latent tree models \cite{Pearl:book}.

\begin{figure}[t]
\subfloat[a][A non-minimal graph]{\label{fig:nonminimal}
\begin{minipage}{1.6in}\centering{
\includegraphics[width=1.1in]{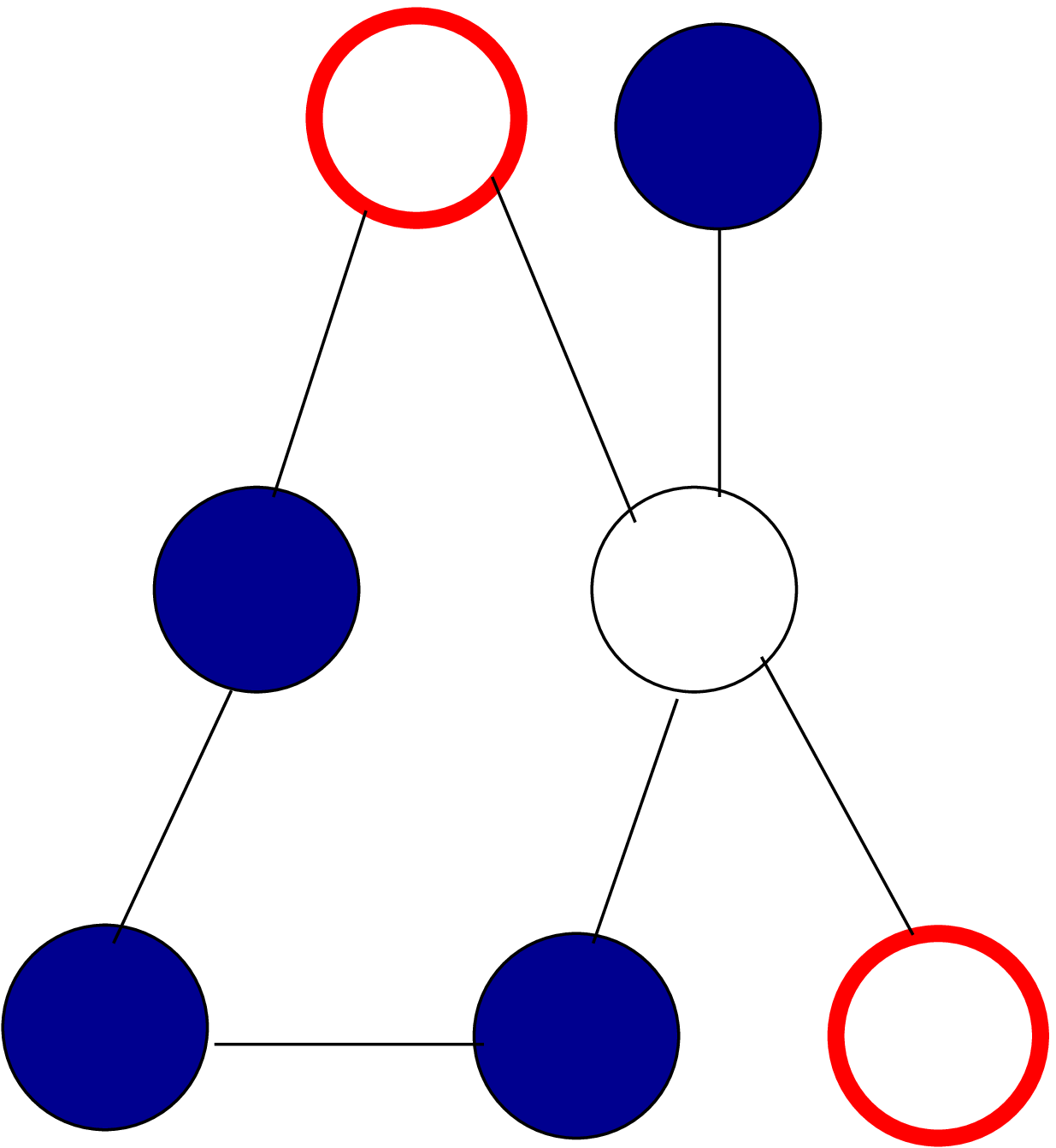}}\end{minipage}}
\hfil\subfloat[b][Minimal representation]{\label{minimal}
\begin{minipage}{1.6in}\centering{
\includegraphics[width=1.1in]{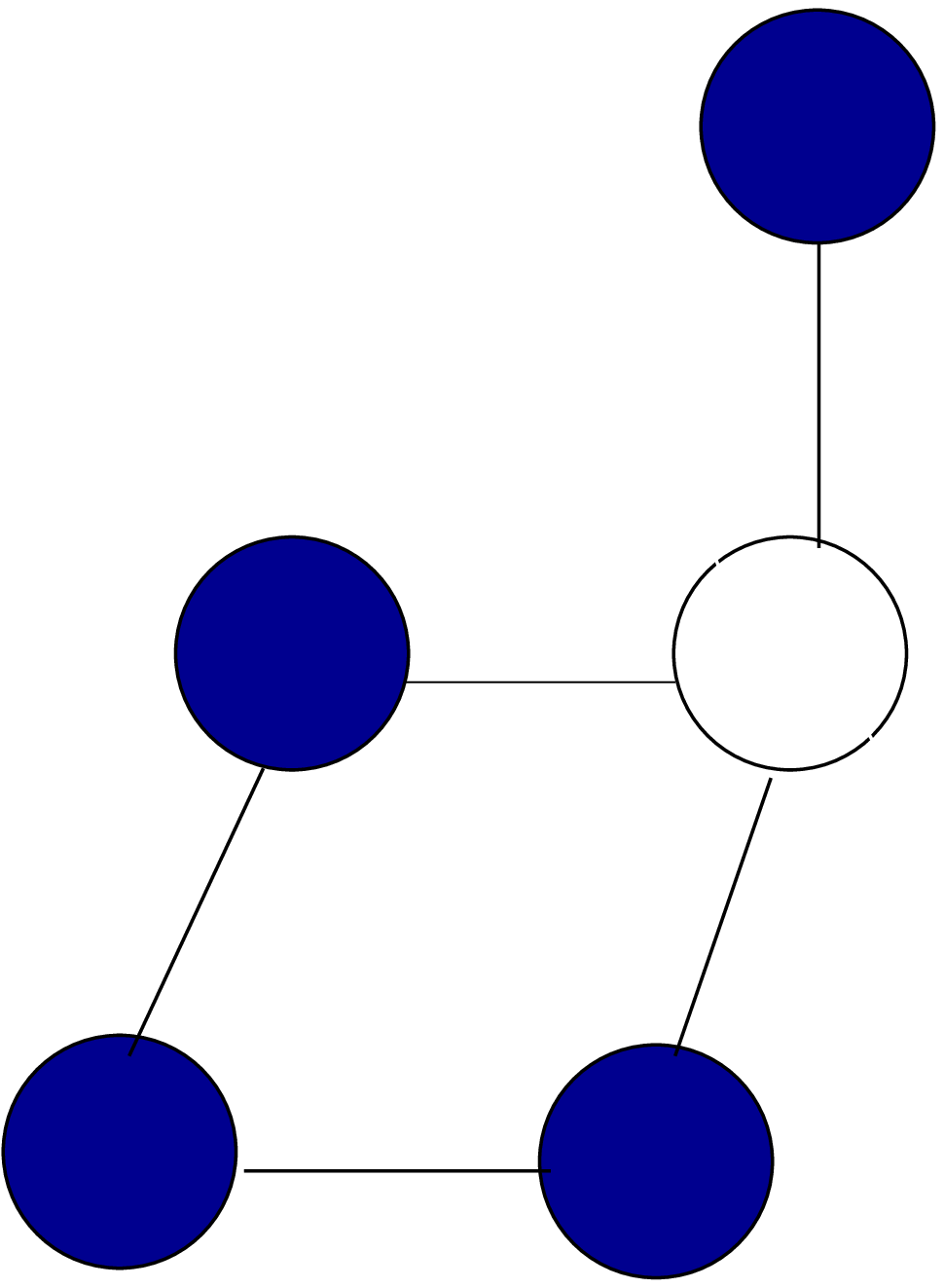}}
\end{minipage}}
\caption{In the above figures, the shaded nodes are participants while the rest are hidden. In the minimal representation of a graph, hidden nodes with degree two and less (the highlighted hidden nodes) are merged with their neighbors. See Procedure~\ref{algo:minimal} for details.}\label{fig:minimal}\end{figure}

There is an equivalence class of topologies with different sets of hidden nodes  which generate the same end-to-end delay distributions between the participating nodes. We refer to the topology with the least number of hidden nodes in this equivalence class as the {\em minimal representation}. Such a minimal representation does not have redundant hidden nodes.
For example, in Fig.\ref{fig:minimal}, the graph and its minimal representation are shown.
In Procedure~\ref{algo:minimal}, we characterize the relationship between a graph and its   minimal representation, given a set of participants. The minimal representation is obtained by iteratively removing redundant hidden nodes (degree two and less) from the graph, i.e., in the first iteration, redundant hidden nodes are removed and the resulting graph is again inspected for the presence of hidden nodes. For example, in Fig.\ref{fig:minimal}, the highlighted hidden nodes are redundant and are thus merged with their neighbors to obtain the minimal representation.


\floatname{algorithm}{Procedure}

\begin{algorithm}[t]\begin{algorithmic}
\STATE Input: Graph $G_{n'}$, set of participating nodes $V_n$, and set of hidden nodes $H_n$.
\STATE Initialize $\tilde{G}_n=G_n$, $n\leftarrow n'$.
\WHILE{$\exists  h \in \tilde{G_n}\cap H_n$ with $\Deg(h)\leq 2$}
\STATE  Remove $h$ from $\tilde{G}_n$ if $\Deg(h)\leq 1$.
\STATE Contract all $h$ with $\Deg(h)= 2$ in $\tilde{G}_n$.
\STATE Decrement $n$ accordingly.
\ENDWHILE
 \end{algorithmic}
\caption{$\tilde{G}_n :=\minimal(G_n;V_n)$ is the minimal representation of $G_{n'}$ given set of participating nodes $V_n$.} \label{algo:minimal}
\end{algorithm}

\floatname{algorithm}{Algorithm}
%

Any algorithm can only reconstruct the unknown topology up to its  minimal representation using  only end-to-end delay information between the participating nodes. In sparse random graphs, only a small (but a linear) number of nodes are removed in the minimal representation, and this number decreases with the average degree $c$. It thus suffices to reconstruct the minimal representation of the original topology, and our goal is to accomplish it using small fraction of participants. We assume that the delay distributions on the edges of the minimal representation $\{D_e\}_{e\in \tilG_n}$ have bounded   variances $\{l(e)\}_{e\in  \tilG_n}$ satisfying
\beq 0<f\leq l(e) \leq g<\infty, \quad \forall\, e\in \tilG_n. \label{eqn:varbound}\eeq

\subsection{Performance Measures}

We now define performance measures for topology discovery algorithms. It is desirable to have an algorithm which outputs a graph structure which is close to the original graph structure. However, the reconstructed graph cannot be directly compared with the original graph since the hidden nodes introduced in the reconstructed graph are unlabeled and may correspond to different hidden nodes in the original graph. To this end, we require the notion of edit distance defined below.
\begin{definition}[Edit Distance]
Let $F,G$ be two graphs\footnote{We consider inexact graph matching where the unlabeled nodes can be unmatched. This is done by adding required number of isolated unlabeled nodes in the other graph, and considering the modified adjacency matrices~\cite{Bunke:83}.} with adjacency matrices $\bfA_F, \bfA_G$,   and let $V $ be the set of labeled vertices in both the graphs (with identical labels). Then the edit distance between $F,G$ is defined as
\[\Delta(F,G;V):=\min_{\pi} || \bfA_F - \pi(\bfA_G)||_1,\]
   where $\pi$ is any permutation on the unlabeled nodes  while keeping the labeled nodes fixed.
\end{definition}
In other words, the edit distance is the minimum number of entries that are different in $\bfA_F$ and in any permutation of $\bfA_G$ over the unlabeled nodes. In our context, the labeled nodes correspond to the participating nodes while the unlabeled nodes correspond to hidden nodes.

Our goal is to output a graph with small edit distance with respect to the minimal representation of the original graph. Ideally, we would like the edit distance to decay as we obtain more delay samples and this is the notion of consistency.

\bd[Consistency]
Denote $\hG_n(\{\bfD^m_{i,j}\}_{i,j\in V_n})$ as the estimated graph using $m$ delay samples between the participating nodes $V_n$. A graph estimator  $\hG_n(\{\bfD^m_{i,j}\}_{i,j\in V_n}) $ is {\em structurally consistent} if  it asymptotically recovers the minimal representation of the unknown topology, i.e.,
\begin{equation}
\lim_{m\to\infty} \Pbb [ \Delta(\hG(\{\bfD^m_{i,j}\}_{i,j\in V_n}),\tilde{G}_n;V_n)>0]=0.
\label{eqn:structural_consistency}
\end{equation}\ed

The above definition assumes that the network size $n$ is fixed while the number of samples $m$ goes to infinity. A more challenging setting where both the network size and the number of samples grow is known as the setting of {\em high-dimensional inference}~\cite{Wainwright&Jordan:08NOW}.
In this setting, we are interested in estimating large network structures using a small number of delay samples.
We will consider this setting for topology discovery in this paper.
Indeed in practice, we have large network structures but can obtain only few end-to-end delay samples with respect to the size of the network. This is formalized using the notion of  sample complexity defined below for our setting.

\bd[Sample Complexity]\label{def:samplecomplexity}If the number of samples is $m =\Omega( f(n))$, for some function $f$, such that the  estimator $\hG_n(\{\bfD^m_{i,j}\}_{i,j\in V_n})$ satisfies
\[\lim_{\substack{m,n\to\infty\\ m=\Omega(f(n))}}\Pbb[\Delta(\hG(\{\bfD^m_{i,j}\}_{i,j\in V_n}),\tilde{G}_n;V_n)=O(g(n))]=0,\] for some function $g(n)$, then the estimator $\hG_n$ is said to have sample complexity of $\Omega(f(n))$ for achieving an edit distance of $O(g(n))$. \ed

Thus, our goal is to discover topology in high-dimensional regime, and design a graph estimator that requires a small number of delay samples, and output a graph with  a small edit distance.

\section{Preliminaries}

We now discuss some simple concepts which will be incorporated into our topology discovery algorithms.

\subsection{Delay Variance Estimation}

In our setting, topology discovery  is based on the end-to-end delays between the participating nodes.
Recall that in Section~\ref{sec:delaymodel}, we assume general delay distributions on the edges  with bounded variances. Our topology discovery algorithms will be based solely on the estimated variances using the end-to-end delay samples.

We use the standard unbiased estimator for variances \cite{Lehmann:book}.
\beq\hl^m(i,j):= \frac{1}{m-1} \sum_{k=1}^m ( D_{i,j}(k)-\bar{D}^m_{i,j})^2, \eeq where $\bar{D}^m_{i,j}$ is the sample mean delay \beq \bar{D}^m_{i,j} := \frac{1}{m}\sum_{k=1}^m D_{i,j}(k).\eeq Note that we do not use an estimator specifically tailored for a parametric delay distribution, and hence, the above estimator yields unbiased estimates for any  delay distribution.

Our proposed algorithms for topology discovery require only the estimated delay variances $\{\hl(i,j)\}_{i,j\in V}$ as inputs. Indeed, more information is available in the delay samples $
\bfD^m$. For instance, in \cite{Bhamidi&etal:09Rand}, the higher-order moments of the delay distribution are estimated using the delay samples and this provides an estimate for the delay distribution. However, we see that for our goal of topology discovery, the estimated end-to-end delay variances suffice and yield good performance.

Recall that $\{l(i,j)\}_{i,j\in V}$ denotes the true end-to-end delay variances and that from   \eqref{eqn:additive},  the variances are additive along any path in the graph.   We will henceforth refer to the variances as ``distances'' between the nodes and the estimated variances as ``estimated distances''. This abstraction also implies that our algorithms will work under input of estimates of  any additive metrics.

\subsection{Quartet Tests}

We first recap the so-called quartet tests, which are building blocks of many algorithms for discovering phylogenetic-tree topologies with hidden nodes~\cite{Ban86, erdos99,Jia01,Pearl:book}. The definition of a quartet is given below. See Fig.\ref{fig:quartet}.

\begin{figure}[t]\centering{\bp\psfrag{a}[l]{$a$}
\psfrag{a}[l]{$a$}\psfrag{h1}[l]{$h_1$}\psfrag{h2}[l]{$h_2$}
\psfrag{b}[l]{$b$}\psfrag{c}[l]{$u$}\psfrag{d}[l]{$v$}
\includegraphics[width=2in]
{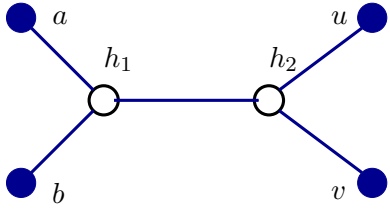}\ep}\caption{Quartet $Q(ab|uv)$. See \eqref{eqn:quartet} and \eqref{eqn:middleedge}.}\label{fig:quartet}\end{figure}

\bd[Quartet or Four-Point Condition]The pairwise distances $\{l(i,j)\}_{i,j\in \{a,b,u,v\}}$  for the configuration in Fig.\ref{fig:quartet} satisfy \beq l(a,u) + l(b,v)= l(b,u) + l(a,v),\label{eqn:quartet}\eeq and the configuration is denoted by   $Q (ab|uv)$. \ed

In the literature on tree reconstruction, instead of \eqref{eqn:quartet}, an inequality test is usually employed since it is more robust, given by,
\beq l(a,b)+ l(u,v) < \min(l(a,u) + l(b,v),l(b,u) + l(a,v))\label{eqn:quartet_inequality}.\eeq However, we use the equality test in \eqref{eqn:quartet}, since it is also useful in detecting cycles present in random graphs.

In practice, we only have access to   distance estimates and we relax the equality constraint  in \eqref{eqn:quartet} to a threshold test, and this  is known as the quartet test.  Thus, the quartet test is local test  between tuples of four nodes.  For the quartet $Q=(ab|uv)$, let $e$ denote the middle edge of the quartet\footnote{Such a middle edge always exists, by allowing for zero length edges, and such trivial edges are contracted later in the algorithm.}, i.e., the edge which joins a vertex on the shortest path between $a$ and $b$ to a vertex on the shortest path between $u$ and $v$ (Note that the edge can have zero length if the hidden nodes connecting $a,b$ and $u,v$ are the same.). The estimated length of the middle edge $(h_1,h_2)$ between hidden nodes $h_1$ and $h_2$  is given by
\beq 2\hl(h_1,h_2) = \hl(a,u) + \hl(b,v) - \hl(a,b)-\hl(u,v). \label{eqn:middleedge}\eeq
Similarly, all other edge lengths of the quartet can be calculated through the set of linear equations which are based on the fact that the end-to-end lengths in a quartet are the sum of edge lengths along the respective paths.

Many phylogenetic-tree reconstruction algorithms proceed by iteratively merging quartets to obtain a tree topology. See~\cite{Durbin:book} for details. We  employ the quartet test for random graph discovery but it additionally incorporates the   presence of cycles. Moreover, we introduce modifications under scenario 2,  as outlined in Section~\ref{sec:scenario2}, where second shortest path distances are available in addition to the shortest path distances between the participating nodes.


\section{Proposed Algorithms}

\subsection{Scenario 1}\label{sec:scene1}

We propose the algorithm $\estone$ for discovering random graphs under scenario 1, as outlined in Section~\ref{sec:scenarios}, where only shortest path distance estimates are available between the participating nodes. The idea behind $\estone$ is similar to the classical phylogenetic-tree reconstruction algorithms based on quartet tests~\cite{Pearl:book,erdos99}. However,   the effect of cycles on such tests needs to analyzed, and is carried out in Section~\ref{sec:cycles}. The algorithm is summarized in Algorithm~\ref{algo:scenario1}.

The algorithm recursively runs the quartet tests over the set of participating nodes. The algorithm limits to   testing only ``short quartets'' between nearby participating nodes. Intuitively, this is done to avoid testing  quartets on short  cycles, since in such scenarios,   the quartet tests may fail to reconstruct the graph accurately.  Since the random graphs are locally-tree like and contain a small number of short cycles,  limiting to short quartets enables us to avoid most of the cycles.  The idea of short quartets has been used before (e.g. in~\cite{erdos99}) but for a different goal of obtaining low sample complexity algorithm for phylogenetic-tree reconstruction. We carry out a detailed analysis  on the effect of cycles on quartet tests in Section~\ref{sec:cycles}.

In algorithm $\estone$, we consider short quartets, where all the estimated distances between   the quartet end points are at most $R g+ \tau$, where $g$ is the upper bound on the (exact) edge lengths in the original graph, as assumed in \eqref{eqn:varbound}. Thus,   $R':=Rg/f$ is the maximum number of  hops between the end points of a short quartet, where $f$ is the lower bound on the edge lengths. We refer to  $R'$ as the {\em diameter of the quartet}. This  needs to be  chosen  carefully to balance the following two events: encountering short cycles and ensuring that most hidden edges (with at least one hidden end point) are part of short quartets.
The parameter $\tau$  is chosen to relax the bound, since we have distance estimates, computed using samples, rather than exact distances between the participating nodes.
The short quartets are listed in arbitrary order in $\Qc$.

The algorithm  attempts to  merge the quartets in $\Qc$, one at a time,  with the previously constructed graph $\hG_n$ using procedure  $\mergequartet$.  There are different possibilities during this process. The quartet under consideration, say $Q(ab|uv)$, may be already satisfied in $\hG_n$: nothing needs to be done in such a scenario; or the quartet may be merged without creating new cycles. This is carried out using procedure $\TreeMerge$. Alternatively, if a cycle needs to be created in $\hG_n$ to merge $Q(ab|uv)$, additional testing needs to be carried out. Firstly, if it is a short cycle (of length less than $2Rg+\tau$), then  the algorithm cannot be guaranteed to merge $Q(ab|uv)$ accurately and it is listed as a bad quartet. Secondly, if it is not a short cycle, the algorithm needs to infer the joining points between the existing paths in $\hG_n$ and the new path to be created. This is carried out using procedure $\CycleMerge$ and entails the presence of ``witnesses'' $\Wc\subset \Qc$, which are (remaining) short quartets whose nodes are within distance $2R$ from $a,b,u,v$. The algorithm attempts to merge the quartets in $\Wc$   without creating new cycles in $\hG_n$, and then attempts to merge $Q(ab|uv)$ by  using existing hidden nodes in the paths to create a new (long) cycle and checking if it conflicts with the distances on quartets in $\Wc$. There is a tolerance of $\epsilon'$ for checking distance conflicts.
In the end, any edge smaller than a threshold $\epsilon$ are contracted, for some chosen constant $\epsilon<f$, where $f$ is the lower bound on the edge lengths of the original graph.

The quartets that fail to be merged using the above procedure are listed as bad quartets. These set of quartets cannot be guaranteed to be merged accurately. Any post-processing heuristic can be used to attempt the merging of these bad quartets. Our analysis accounts for these bad quartets towards contributing to the edit distance between the reconstructed graph and the minimal representation of the original graph. The above algorithm is similar in spirit to quartet merging algorithm proposed in~\cite{erdos99}, but with the crucial addition of $\CycleMerge$ procedure to handle the presence of cycles.




\begin{algorithm}[t]\begin{algorithmic}
\STATE Input: Distance estimates between the participating nodes $\{\hl(i,j)\}_{i,j\in V_n}$, upper bound $g$ on exact edge lengths and parameters $R, \tau,\epsilon, \epsilon'>0$.
\STATE Initialize list of short quartets: $\Qc=\{ Q(ab|uv):
\max\limits_{i,j\in \{a,b,u,v\}}\hl(i,j) < Rg +\tau\}$, list of bad quartets $\Qc_{\bad}=\emptyset$ and reconstructed graph $\hG_n =(V_n, \emptyset)$.
\WHILE{$\Qc\neq \emptyset$}
\STATE $Q(ab|uv)\leftarrow \mbox{Pop}(\Qc)$.
\STATE $\Qc\leftarrow \Qc\setminus \{Q(ab|uv)\}$.
\STATE $[\hG_n,\mbox{Fail}]\leftarrow \mergequartet(\hG_n, Q(ab|uv), \Qc,\{\hl(i,j)\}_{i,j\in V_n};\epsilon, \epsilon')$.
\IF{$\mbox{Fail}$}\STATE $\Qc_{\bad}\leftarrow \Qc_{\bad}\cup \{Q(ab|uv)\}$.
\ENDIF
\ENDWHILE
\STATE Use any heuristic to merge bad quartets in $\Qc_{\bad}$.
 \end{algorithmic}
\caption{$\estone(\{\hl(i,j)\}_{i,j\in V_n};R,g,\tau,\epsilon,\epsilon')$ for Topology Discovery Using Shortest-Path Distance Estimates.} \label{algo:scenario1}
\end{algorithm}

\floatname{algorithm}{Algorithm}

\begin{algorithm}[t]\begin{algorithmic}
\STATE Input: Shortest-path   and second shortest-path distance estimates   $\{\hl(i,j),\hl_2(i,j)\}_{i,j\in V_n}$ upper bound $g$ on exact edge lengths and parameters $R, \tau,\epsilon, \epsilon'>0$.
\STATE Initialize list of short quartets: $\Qc=\{ Q(ab|uv):
\max\limits_{i,j\in \{a,b,u,v\}}\hl(i,j) < Rg +\tau\}$, list of bad quartets $\Qc_{\bad}=\emptyset$ and reconstructed graph $\hG_n =(V_n, \emptyset)$.
\WHILE{$\Qc\neq \emptyset$}
\STATE $Q(ab|uv)\leftarrow \mbox{Pop}(\Qc)$.
\STATE $\Qc\leftarrow \Qc\setminus \{Q(ab|uv)\}$.
\STATE $[\hG_n,\mbox{Fail}]\leftarrow \mergequartet(\hG_n, Q(ab|uv), \Qc,\{\hl(i,j),\hl_2(i,j)\}_{i,j\in V_n};\epsilon, \epsilon')$.
\IF{$\mbox{Fail}$}\STATE $\Qc_{\bad}\leftarrow \Qc_{\bad}\cup \{Q(ab|uv)\}$.
\ENDIF
\ENDWHILE
\STATE Use any heuristic to merge bad quartets in $\Qc_{\bad}$.
 \end{algorithmic}
\caption{$\esttwo(\{\hl(i,j),\hl_2(i,j)\}_{i,j\in V_n};R,g,\tau,\epsilon,\epsilon')$ for Topology Discovery Using Shortest-Path and Second Shortest-Path Distance Estimates.} \label{algo:scenario2}
\end{algorithm}

\floatname{algorithm}{Procedure}

\begin{algorithm}[t]\begin{algorithmic}
\STATE Input: Current graph $\hG$, candidate  quartet $Q(ab|uv)$, remaining short quartets $\Qc$, shortest  distance estimates between the participating nodes $\{\hl(i,j)\}_{i,j\in V_n}$ and optionally second shortest distances $\{\hl_2(i,j)\}_{i,j\in V_n}$, threshold $\epsilon$ for contracting short edges and tolerance $\epsilon'$ for comparing path lengths.
\IF{For each $i,j\in \{a,b,u,v\}$, $|\hl(i,j;\hG)-\hl(i,j;Q)|< \epsilon'$ (All paths already present in $\hG$)}
\STATE $\mbox{Fail}\leftarrow \mbox{False}$.
\ELSIF{For each $i,j\in \{a,b,u,v\}$, either $|\hl(i,j;\hG)-\hl(i,j;Q)|< \epsilon'$ or $\hl(i,j;\hG)=\infty$ (Either the paths agree or  the path  does not exist in $\hG$)}
\STATE $\hG_n \leftarrow \TreeMerge(\hG_n, Q(ab|uv),\epsilon')$, $\mbox{Fail}\leftarrow \mbox{False}$. (Merge quartet without creating cycles).
\ELSIF{$\exists\,i,j\in \{a,b,u,v\}$ such that $\hl(i,j;\hG)+ \hl(i,j;Q) <2Rg +\tau$  (A merge would create a short cycle)}
\IF{Second shortest distances $\{\hl_2(i,j)\}_{i,j\in V_n}$, are available}
 \STATE Use second shortest distances between $a,b,u,v$ to infer the join points in $\hG$. If the points are consistently found, add quartet $Q(ab|uv)$ to $\hG$ and output $\mbox{Fail}\leftarrow\mbox{False}$. Else output $\mbox{Fail}\leftarrow\mbox{True}$. (Report failure due to inconsistent distances).
\ELSE \STATE $\mbox{Fail}\leftarrow\mbox{True}$.  (Report failure due to presence of a short cycle).
\ENDIF\ELSE
\STATE $[\hG_n,\mbox{Fail}] \leftarrow \CycleMerge(\hG_n, Q(ab|uv),\Qc,\{\hl(i,j),[\hl_2(i,j)]\}_{i,j\in V_n},[\{\hl_2(i,j)\}_{i,j\in V_n}];\epsilon')$. (Attempt to merge quartet by creating a new long cycle and querying witnesses, if no witnesses are present, create a new path, else  if witnesses are  contradictory output fail).
\ENDIF
\STATE Contract any edge (with at least one hidden end point) with length $< \epsilon$.
 \end{algorithmic}
\caption{$\mergequartet(\hG_n, Q(ab|uv)), \Qc, \{\hl(i,j),[\hl_2(i,j)]\}_{i,j\in V_n};\epsilon, \epsilon')$ for merging a new quartet $Q(ab|uv)$ with current structure $\hG_n$.} \label{algo:mergequartet}
\end{algorithm}

\begin{algorithm}[t]\begin{algorithmic}
\STATE Input: Current graph $\hG$, candidate  quartet $Q(ab|uv)$ with hidden nodes $h_1, h_2$ (See Fig.\ref{fig:quartet}),   and tolerance $\epsilon'$ for comparing path lengths.
\IF{There exists hidden node in $\hG$ such that $|\hl(i,h;\hG)-\hl(i,h_1;Q)|<\epsilon'$ for $i=a,b$}
\STATE Assign $h$ as $h_1$.
\ENDIF
\IF{There exists hidden node in $\hG$ such that $|\hl(i,h;\hG)-\hl(i,h_2;Q)|<\epsilon'$ for $i=u,v$}
\STATE Assign $h$ as $h_2$.
\ENDIF
\STATE Connect paths in $\hG$ according to $Q$ which are missing as follows:
\STATE If both $h_1$ and $h_2$ are assigned in $\hG$,  connect $h_1$ and $h_2$ in $\hG$ and assign length $\hl(h_1,h_2;Q)$ if they are not already connected in $\hG$.
\STATE If say $h_1$ is assigned and the path between $a$ and $h_1$ exists in $\hG$ but not between $b$ and $h_1$, let $l\leftarrow \min_{w} 0.5(\hl(w,b) + \hl(h_1,b;Q)- \hl(h_1,w;\hG)$ over all $w\in \hG$ such that $\hl(w,b)< Rg+\tau$. If $l < \hl(h_1, b;Q)$, split path $\hl(h_1,w;\hG)$, add new hidden node $h_3$ such that $\hl(h_3,w;\hG)=\hl(b,w)-l $ and attach $b$ to $h_3$ with length $l$; otherwise create a new path between $h_1$ and $b$. Similarly split the other paths if present or create new paths.
\STATE If the paths exist but not the hidden nodes, split the paths and add hidden nodes according to the lengths in $Q$.  Otherwise, also add new paths to $\hG$.
 \end{algorithmic}
\caption{$\TreeMerge(\hG_n, Q(ab|uv));\epsilon')$ for merging a new quartet with current structure $\hG_n$ without creating cycles.} \label{algo:TreeMerge}
\end{algorithm}

\begin{algorithm}[t]\begin{algorithmic}
\STATE Input: Current graph $\hG$, candidate  quartet $Q(ab|uv)$ with hidden nodes $h_1, h_2$ (See Fig.\ref{fig:quartet}), remaining short quartets $\Qc$, shortest distance estimates $\{\hl(i,j)\}_{i,j\in V_n}$ and optionally second shortest distance estimates $\{\hl_2(i,j)\}_{i,j\in V_n}$ and tolerance $\epsilon'$ for comparing path lengths.
\STATE $\Wc\leftarrow \{Q(ij|kl): Q(ij|kl)\in \Qc, \cup_{x,y \in \{i,j,k,l,a,b,u,v\}}\hl(x,y)< 2Rg+\tau\}$. (Find potential witnesses by using short quartets ``near'' to $a,b,u,v$. Also use second shortest distances if available and they are less than $2Rg+\tau$).
\FOR{Each $Q(wx|yz)\in \Wc$}
\IF{For   $i,j\in \{w,x,y,z\}$, either $|\hl(i,j;\hG)-\hl(i,j;Q)|< \epsilon'$ or $\hl(i,j;\hG)=\infty$ (Either the paths agree in the two graphs or  the path  does not exist in $\hG$)}
\STATE $\hG_n \leftarrow \TreeMerge(\hG_n, Q(wx|yz),\epsilon')$, $\mbox{Fail}\leftarrow \mbox{False}$. (Merge all quartets in $\Wc$ which do not create   cycles).
\ENDIF
\ENDFOR
\STATE To create new paths in $\hG$ according to $Q(ab|uv)$, consider all hidden nodes on paths in $\hG$ as candidates for positions where the paths split. Query the quartet corresponding to these hidden nodes for verification. (See Fig.\ref{fig:quartetverify} for an example).
\STATE If the witnesses are absent or contradictory, output $\mbox{Fail}\leftarrow \mbox{True}$. Else, add the new path to $\hG$ and output $\mbox{Fail}\leftarrow \mbox{False}$.
 \end{algorithmic}
\caption{$\CycleMerge(\hG_n, Q(ab|uv)), \Qc, \{\hl(i,j)\}_{i,j\in V_n};\epsilon')$ for merging a new quartet with current structure $\hG_n$ by creating cycles.} \label{algo:CycleMerge}
\end{algorithm}

%

%

\subsection{Scenario 2}\label{sec:scenario2}

We now consider scenario 2, as outlined in Section~\ref{sec:scenarios}, where second shortest path distance estimates are available in addition to shortest path distance estimates between the participating nodes. We propose $\esttwo$ algorithm for this case, which is summarized in Algorithm~\ref{algo:scenario2}.

The algorithm $\esttwo$ is an extension of $\estone$, where we use the second shortest distances in the quartet tests, in addition to the shortest distances. For each tuple of participating nodes $a,b,u,v\in V_n$, the quartet test in \eqref{eqn:quartet} is carried out for all possible combinations of shortest and second shortest distances; only short quartets are retained, where all the distances used for quartet test are less than the specified threshold (which is the same as in $\estone$). If the same quartet is formed using different combinations of shortest and second shortest distances, only the quartet with the shorter middle edge, computed using \eqref{eqn:middleedge}, is retained. We clarify the reason behind this rule and give examples on when this can occur in Section~\ref{sec:cycles}.
As before, all these quartets are merged with previously constructed graph using procedure $\mergequartet$, but with a minor difference that the path lengths need to be checked since there may  be multiple paths between participating nodes with different lengths. The performance analysis for $\esttwo$ is carried out in Section~\ref{sec:scenario2_analysis}.

\section{Analysis  Under Exact Distances}

We now undertake performance analysis for the proposed topology discovery algorithms $\estone$ and $\esttwo$. In this section, for simplicity,  we first analyze the performance assuming that exact distances between the participating nodes are input to the algorithms.   Analysis when distance estimates are input to the algorithms is considered in Section~\ref{sec:samples}.

\subsection{Effect of Cycles on Quartet Tests}\label{sec:cycles}

\begin{figure}[t]\centering{\bp\psfrag{a}[l]{$a$}
\psfrag{a}[l]{$a$}\psfrag{h1}[l]{$h_1$}\psfrag{h2}[l]{$h_2$}
\psfrag{b}[l]{$b$}\psfrag{c}[l]{$u$}\psfrag{d}[l]{$v$}
\psfrag{w}[l]{$w$}\psfrag{x}[l]{$x$}\psfrag{y}[l]{$y$}
\psfrag{z}[l]{$z$}\psfrag{h3}[l]{$h_3$}\psfrag{h4}[c]{$h_4$}
\includegraphics[width=2in]
{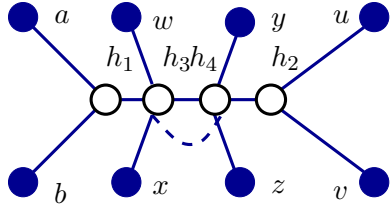}\ep}\caption{An example on the use of witnesses  for creating new paths in the existing graph $\hG$ in $\CycleMerge$ Procedure. In order to create a new path between $h_1$ and $h_2$ (shown using dotted lines according to length specified by quartet $Q(ab|uv)$), the quartet $Q(wx|yz)$ is used as a witness to verify if $h_3$ and $h_4$ are the joining points of the existing path in $\hG$ with  the new path. }\label{fig:quartetverify}\end{figure}

\begin{figure}[t]\centering{\bp\psfrag{a}[l]{$a$}
\psfrag{a}[l]{$a$}\psfrag{h1}[l]{$h_1$}\psfrag{h2}[l]{$h_2$}
\psfrag{b}[l]{$b$}\psfrag{c}[l]{$u$}\psfrag{d}[l]{$v$}
\includegraphics[width=2in]
{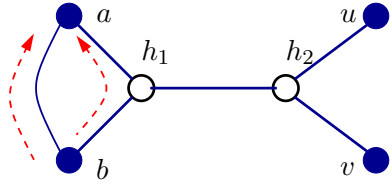}\ep}\caption{A bad quartet: $l(a,b)< l(a,h_1) + l(b,h_1)$. Since the maximum number of hops between $\{a,b,u,v\}$ is $R':=Rg/f$, and one of the shortest paths is not along the quartet, the middle edge $(h_1,h_2)$ is part of a generalized cycle of (hop) length less than $2R'$. Such quartets are detected by the $\estone$ algorithm.}\label{fig:badquartet}\end{figure}

\begin{figure*}[t]\subfloat[a][Bad Quartet Where Test Fails.]{\begin{minipage}{2in}
\centering{\label{fig:badquarteta}
\bp\psfrag{a}[l]{$a$}
\psfrag{a}[l]{$a$}\psfrag{h1}[l]{$h_1$}\psfrag{h2}[l]{$h_2$}
\psfrag{b}[l]{$b$}\psfrag{c}[l]{$u$}\psfrag{d}[l]{$v$}
\includegraphics[width=2in]
{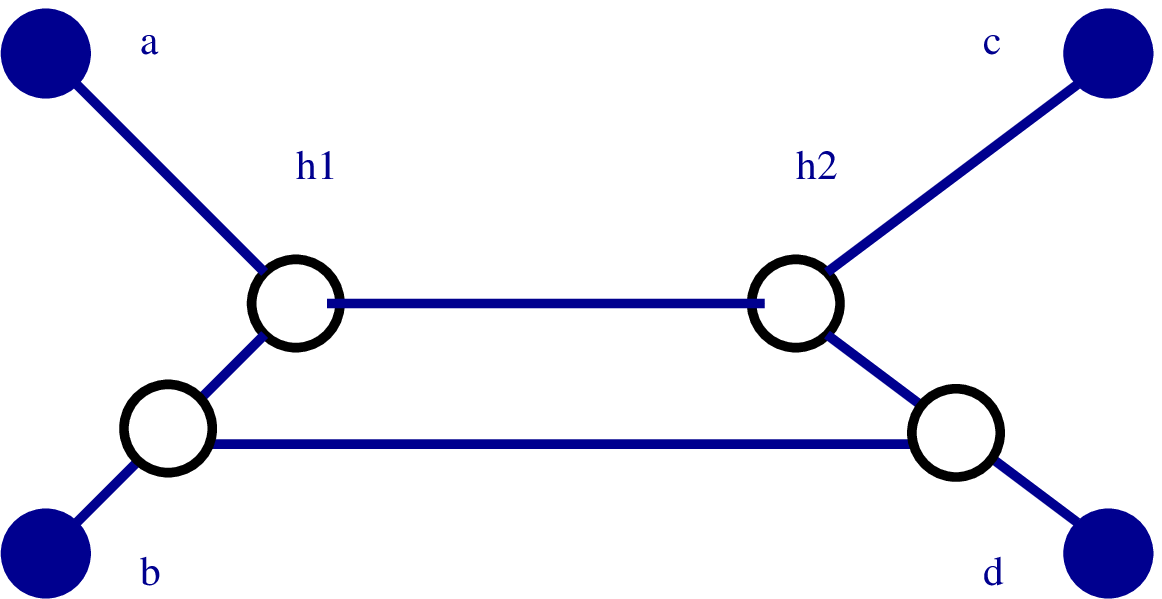}\ep}\end{minipage}}\hfil
\subfloat[b][Bad Quartet Where Test Succeeds.]{\label{fig:badquartetb}\begin{minipage}{2in}\centering{\bp\psfrag{a}[l]{$a$}
\psfrag{a}[l]{$a$}\psfrag{h1}[l]{$h_1$}\psfrag{h2}[l]{$h_2$}
\psfrag{b}[l]{$b$}\psfrag{c}[l]{$u$}\psfrag{d}[l]{$v$}
\includegraphics[width=2in]
{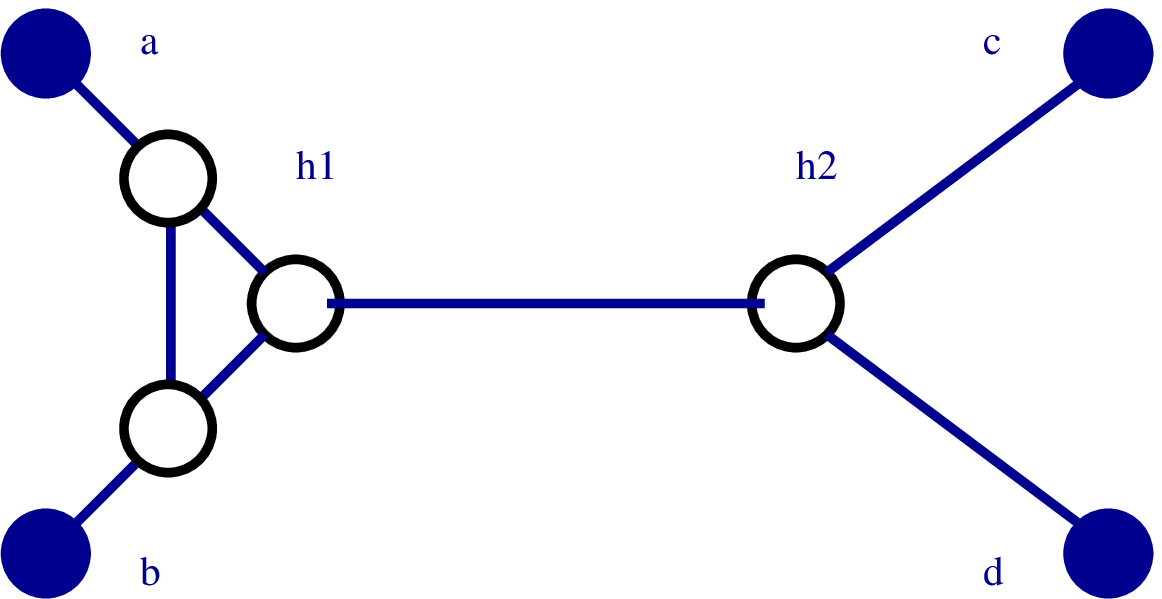}\ep}\end{minipage}}\hfil
\subfloat[c][Reconstructed Quartet for (b).]{\label{fig:badquartetc}\begin{minipage}{2in}\centering{\bp\psfrag{a}[l]{$a$}
\psfrag{a}[l]{$a$}\psfrag{h1}[l]{$h_1$}\psfrag{h2}[l]{$h_2$}
\psfrag{b}[l]{$b$}\psfrag{c}[l]{$u$}\psfrag{d}[l]{$v$}
\psfrag{1.5}[l]{$1.5$}\psfrag{1}[l]{$1$}
\includegraphics[width=2in]
{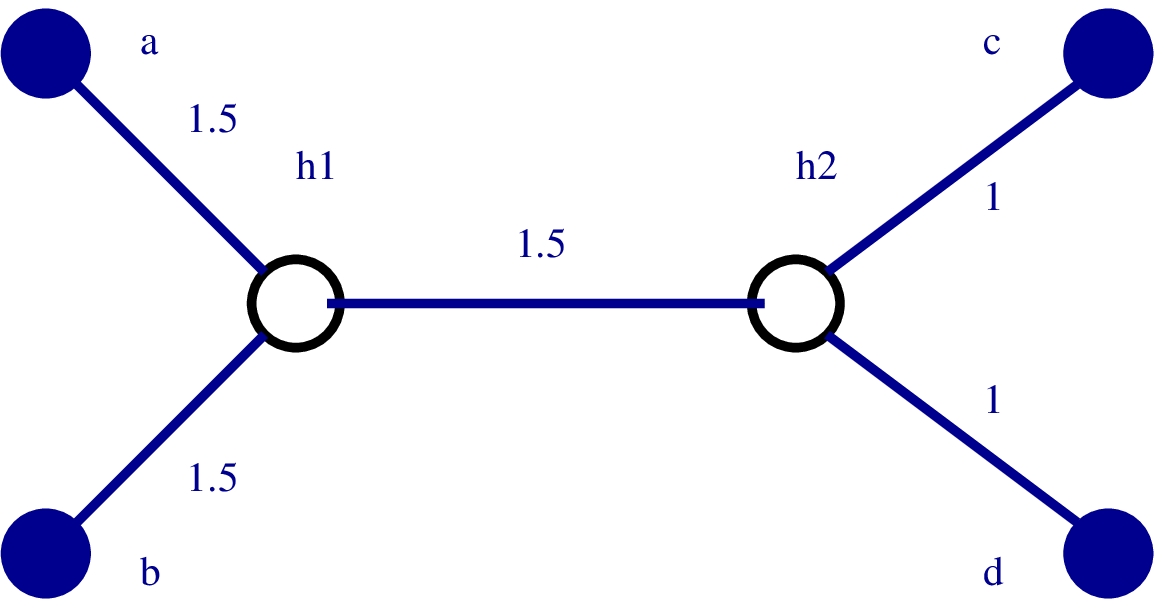}\ep}\end{minipage}}\caption{Two possible outcomes for bad quartets in (a)
 and (b). Assume  all unit-length edges. In (a), the procedure $\mergequartet$ fails and quartet is not declared, while in (b), it succeeds but leads to wrong edge estimates, as shown in (c).}\label{fig:badquartetexample}\end{figure*}

We now analyze the effect of cycles on quartet tests.  Recall that the quartet test is the   inequality test in \eqref{eqn:quartet}, and if this inequality test is satisfied,  internal edge lengths of the quartet are  computed, and they are added to the output using procedure $\mergequartet$.   The quartet test in \eqref{eqn:quartet} is based on the assumption that the shortest paths between the four nodes $\{a,b,u,v\}$ in the quartet are along the paths on the quartet.

Thus,   the outcome of the quartet test is incorrect only when   some shortest path between $\{a,b,u,v\}$ is outside the quartet.    We refer to such quartets as ``bad quartets''. There are two possible outcomes for bad quartets (a) the procedure $\mergequartet$ detects inconsistencies in the set of linear equations, used to compute the internal distances in the quartet, and does not  merge the quartet, or (b) the procedure $\mergequartet$ does not detect inconsistencies, and thus merges a fake quartet with wrong internal edge lengths.  Both these outcomes   result in reconstruction error.

The   examples of both the cases are given in Fig.\ref{fig:badquartetexample}.  Note that the set of linear equations used by the procedure $\mergequartet$ for computing the internal edge-lengths in the quartet consist of $5$ variables and $6$ equations (corresponding to the $6$ known edge-lengths between the quartet end-points). Additionally, there is an equality constraint that $l(a,u) + l(b,v) = l(a,v)+l(b,u)$. The case in Fig.\ref{fig:badquarteta} does not satisfy this equality constraint\footnote{There exist pathological cases of equal distances where configurations of the form in Fig.\ref{fig:badquarteta} will satisfy equality constraint.   Such scenarios  do not occur  in a.e. random graph.}, since the cycle is in the middle of the quartet, and thus the procedure $\mergequartet$ does not merge this quartet. On the other hand, for the case in Fig.\ref{fig:badquartetb}, the equality constraint is satisfied, since the cycle is on the same side of the quartet, and in this case, the procedure $\mergequartet$ merges the quartet, but with wrong edge lengths, as shown in Fig.\ref{fig:badquartetc}.

Thus,   bad quartets lead to reconstruction error.
The   number of bad quartets can be bounded as follows: in a bad quartet,   the middle edge of the quartet is part of  a (generalized) cycle of length less than $2R'$, where $R':=Rg/f$ is the maximum number of hops between the endpoints of a short quartet, as discussed in Section~\ref{sec:scene1}. In addition, the bad quartets also affect the merging of quartets using $\CycleMerge$ procedure when they are called upon to serve as witnesses. Thus, we also need to   consider quartets which are part of slightly longer cycles. See Appendix~\ref{proof:estone} for details.
The number of  such bad quartets    can be bounded for   random graphs leading to reconstruction guarantees for $\estone$ algorithm.

%

For the $\esttwo$ algorithm where second shortest path distances are additionally available,  bad quartets do not adversely affect  performance. We argue that a quartet is correctly recognized as long as  the paths on the quartet correspond to either the shortest or the second shortest paths (between the quartet endpoints). In such a scenario,  some combination of shortest and second shortest path distances exists which accurately reconstructs the quartet and the $\esttwo$ algorithm finds all such   combinations.
Moreover, fake quartets are detected since they produce a longer middle edge than the true quartet. This is because
the cycle shortens the distance between end points on its side (in Fig.\ref{fig:badquartetb}, this corresponds to $\{a,b\}$ and note that the middle edge in Fig.\ref{fig:badquartetc} is longer than the true edge length).


Thus, a quartet is correctly reconstructed under $\esttwo$  when  the paths on the quartet consist of shortest or second shortest paths. We finally use the locally tree-like property of random graphs to establish that this occurs in almost every graph if  the quartet diameter $R'$ is small enough. Thus, we  obtain stronger reconstruction guarantees for $\esttwo$ algorithm.


\subsection{Analysis of $\estone$}\label{sec:scenario1_analysis}

We now provide edit distance guarantees for $\estone$ under appropriate choice of maximum quartet diameter $R':=Rg/f$. We analyze the edit distance by counting the number of  hidden edges (with at least one hidden end point) which are not recovered correctly under $\estone$. A hidden edge is not recovered when one of the following two events occur: (a) it is not part of a short quartet (b) it is part of a bad short quartet. A large value of the quartet diameter $R'$ decreases the likelihood of event (a), while it increases the likelihood of event (b), i.e., we are likely to encounter more cycles as $R'$ is increased. For a fixed value of $R'$, we analyze the likelihood of these two events and obtain the bound on edit distance stated below.

Assume that the algorithm $\estone$ chooses parameter $R$ as \beq\label{eqn:Rupperbound} R_{\min}\leq R\leq R_{\max},\eeq  where \beq R_{\min}:= 2 \frac{ \log \frac{9 \log n}{(\sqrt{c}-1)^2}}{\log 3}, \quad  R_{\max}:=\frac{6 \log n}{5 \log c}\label{eqn:Rmax}.\eeq
Let the fraction of participating nodes be $\rho_n=n^{-\beta}$, such that \beq \rho_n c^{\frac{R-R_{\min}}{2}} = \omega(1),\label{eqn:fracconstraint}\eeq implying that $\gamma> 2 \beta $, where \beq \gamma:=\frac{\log c}{\log n}(R-R_{\min}).\eeq Similarly,   define $\mu$ as
\beq \mu:= R \frac{\log \frac{c}{\xi(c)}}{\log n},\eeq where  $\xi(c)$  is a function that depends on the average degree $c$ of the original Erd\H{o}s-R\'{e}nyi  random graph, and is given by   \beq \xi(c):=1- e^{-c} - c e^{-c} - 0.5 c^2 e^{-c}.\label{eqn:xi}\eeq Recall that $f$ and $g$ are the bounds on edge lengths according to \eqref{eqn:varbound}.
We have the following result.

\bt[Edit Distance Under $\estone$]\label{thm:estone}
The algorithm $\estone$  recovers the minimal representation $\tilG_n$ of the giant component of  a.e. graph $G_n \sim \Gmsc(n, c/n)$ with edit distance
\beq \Delta(\hG_n,\tilG_n;V_n)=  \tilde{O}(n^{5\mu g/f-4\beta}).\label{eqn:editestone}\eeq   \et


\noindent{\bf Remarks: }

\noindent{\em (i) }Thus, an edit-distance guarantee can be provided under $\estone$ when the parameter $R$ is chosen according to the constraints mentioned above. A sufficient condition to achieve a sub-linear edit distance above under homogeneous edge lengths $(f=g)$ is when
\beq   10 \beta(1+\delta) \frac{\log\frac{c}{\xi(c)}}{\log c}- 4\beta < 1,\eeq for some constant $\delta>0$. When $c\to \infty$, we have $\xi(c) \to 1$ and in this regime, we have that $\beta < 1/6$. In other words, approximately $n^{\frac{5}{6}}$ nodes need to participate to achieve a sub-linear edit distance under $\estone$.





\noindent{\em (ii) }When the ratio of the bounds on the edge lengths $g/f$ is small (i.e., the edge lengths are nearly homogeneous), the edit-distance guarantee in \eqref{eqn:editestone} improves, for a fixed  $\rho$. This is because we can control the hop lengths of the selected quartets more effectively in this case.

\noindent{\em (iii) }The dominant event leading to the edit-distance bound in \eqref{eqn:editestone} is  the presence of bad quartets due to  short cycles in the random graph. In subsequent section, we show that $\esttwo$ algorithm effectively handles this event using the second shortest path distances.


\noindent{\bf Proof Ideas: }

The proof is based on the error events that can cause the quartet tests to fail. The first error event is that  an edge which does not occur as a middle edge of a short quartet, meaning that there are not enough participating nodes within distance $R/2$ from it. The second error event is that an edge occurs as a middle edge of a bad quartet, meaning that it is close to a short cycle or it has bad quartets as witnesses. We analyze the probability of these events and the resulting edit distance due to these events.

\subsection{Analysis of $\esttwo$}\label{sec:scenario2_analysis}

We now provide edit distance guarantees for $\esttwo$ algorithm. The analysis is on the lines of the previous section, but we instead analyze the presence of overlapping cycles, as noted in Section~\ref{sec:cycles}.  There are no overlapping short cycles in a random graph, and thus, we can provide a much stronger reconstruction guarantee for the $\esttwo$ algorithm, compared to the $\estone$ algorithm.
 We have the following result.

\bt[Edit Distance Under $\esttwo$]\label{thm:esttwo}
Under the assumptions of Theorem~\ref{thm:estone}, the algorithm $\esttwo$  recovers the minimal representation $\tilG_n$ of the giant component of a.e. graph $G_n \sim \Gmsc(n, c/n)$ with edit distance
\beq \Delta(\hG_n,\tilG_n;V_n)=\tilde{O}(n^{8\mu g/f-4\beta-1} ).\eeq \et

The above result immediately implies that consistent recovery of the minimal representation is possible when there are enough number of participating nodes. We state the result formally below.

\begin{corollary}[Consistency Under $\esttwo$]
The algorithm $\esttwo$ consistently recovers the minimal representation $\tilde{G}_n$ of the giant component of a.e. graph $G_n \sim \Gmsc(n, c/n)$, when the parameter $R$ and the fraction of participating nodes $\rho$ satisfy  \[ \left(\frac{c}{\xi(c)}\right)^{\frac{8Rg}{f}}\rho^4  = o(n),\quad c^{\frac{R-R_{\min}}{2}} \rho =\omega(1),\] or equivalently \[\frac{8\mu g}{f}-4 \beta < 1,\quad \gamma > 2\beta.\]
\end{corollary}

\noindent{\bf Remarks: }

\noindent{\em (i) }From the above constraints, we see that  consistent topology recovery is feasible. Thus, for homogeneous edge lengths $(f=g)$, as $c\to \infty$ and the number of participants is  more than $n^{11/12}$,  $\esttwo$ consistently recovers the topology.  Thus, a sub-linear number of participants suffice to recover the minimal representation consistently.


\noindent{\em (ii) }Thus, the availability of second shortest distances makes consistent topology discovery possible with a sub-linear number of participating nodes, while consistent recovery is not tractable under $\estone$  using only shortest-path distances between a sub-linear number of participants.




\noindent{\bf Proof Ideas: }

The proof is on similar lines as in Theorem~\ref{thm:estone}, but with modified error events that cause the quartet tests to fail. As before,  the first error event is that  an edge which does not occur as a middle edge of a short quartet. The second error event is now that an edge is close to two overlapping short cycles instead of being   close to a single short cycle. This event does not occur in random graphs for sufficiently short lengths, and thus, we see a drastic improvement in edit distance.

\section{Analysis  Under Samples}\label{sec:samples}

We have so far analyzed the performance of $\estone$ and $\esttwo$ algorithms when exact distances (i.e., delay variances) are input to the algorithm. We now analyze the scenario when instead only delay samples are available and estimated variances are input to the algorithm.

We show that the proposed algorithms have low sample complexity, meaning they require slow scaling of number of samples compared to the network size to achieved guaranteed performance. The result is given below.

\bt[Sample Complexity]\label{thm:sample}The edit distance guarantees under $\estone$ and $\esttwo$ algorithms, as stated in Theorem~\ref{thm:estone} and Theorem~\ref{thm:esttwo}, are achieved under input of estimated delay variances, if the number of delay samples satisfies \beq m = \Omega(\poly(\log n)).\eeq \et

Thus, the sample complexity of $\estone$ and $\esttwo$ algorithms is $\poly(\log n)$. In other words, the size of the network $n$ can grow much faster than the number of delay samples $m$, and we can still obtain good estimates of the network. This implies with $m=\Omega(\poly(\log n))$ samples, we can consistently discover the topology under $\esttwo$ algorithm, given sufficient fraction of participating nodes.

\noindent{\bf Proof Ideas: }

The proof follows from  Azuma-Hoeffding  inequality for concentration of individual variance estimates, as in \cite[Proposition 1]{Bhamidi&etal:09Rand}, and then consider the union bound over various events.

\section{Converse Results \& Discussion}

\subsection{Fraction of Participating Nodes}

We have so far provided edit distance guarantees for the proposed topology discovery algorithms.
In this section, we provide a lower bound on the fraction of participating nodes required for any algorithm to recover the original graph up to a certain edit distance guarantee.

We can obtain a meaningful lower bound only when the specified edit distance   is lower than the edit distance between a given graph and an  independent realizations of the random graph. Otherwise, the edit distance guarantee could be realized by a random construction of the output graph. To this end, we first prove a lower bound on the edit distance between any fixed graph and an   independent realization of the random graph.


Let $\Dc(G;\delta)$ denote the set of all graphs which have edit distance of at most $\delta$ from  $G$
\beq \Dc(G;\delta):=\{F: \Delta(F,G;\emptyset)< \delta\}\label{eqn:Dc_def}.\eeq

\begin{lemma}[Lower Bound on Edit Distance]\label{lemma:rand-graphs}Almost every random graph $G_n \sim \Gmsc(n,c/n)$ has an edit distance at least   $(0.5c-1) n$ from any given graph $F_n$.
\end{lemma}
\bprf   First, we have  for any graph $F_n$ \beq\label{eqn:num_graphs_edit_distance}|\Dc(F_n;\delta n)| \le n! \cdot {\frac{n^2}{2} \choose \delta n} < n^{( \delta +1)n}3^{\delta n},\eeq since we can permute the $n$ vertices and change at most $\delta n$ entries in the adjacency matrix $\bfA_F$ and we use the bound that ${N \choose k }\leq \frac{N^k}{k!} \leq  (\frac{N}{k})^k 3^k$.
Let $\Bc$ denote the set of    graphs   having exactly $\frac{cn}{2}$ edges and the size of $\Bc$ is
\[|\Bc| =  {\frac{n^2}{2} \choose \frac{cn}{2}} \ge (\frac{n^2}{cn})^{cn/2} = (\frac{n}{c})^{cn/2}.\]
We can now bound the probability that   a random graph $G_n \sim \Gmsc(n, c/n)$ belongs to set $\Dc(F_n;\delta n)$ for any given graph $F_n$ is
\begin{align}\nn \Pbb[ G_n\in \Dc(F_n;\delta n)] &\leq \frac{\Pbb[G_n\in \Dc(F_n;\delta n)]}{\Pbb[G_n\in \Bc]}\\ &\leq \frac{|\Dc(F_n;\delta n)| \max\limits_{g\in \Dc(F_n;\delta n)} \Pbb[G_n=g]}{|\Bc|\min\limits_{g\in \Bc}\Pbb[G_n=g]}\nn \\ &\overset{(a)}{\leq} \frac{|\Dc(F_n;\delta n)| }{|\Bc| }= n^{(\delta+1 - c/2)n}3^{\delta n}\nn, \end{align} where inequality (a) is due to the fact that $\min_{g\in \Bc}\Pbb[G_n=g]\geq \max_{g\in \Sc(F_n)}\Pbb[G_n=g]$ (i.e., the mode of the binomial distribution). Hence, $\Pbb[ G_n\in \Dc(F_n;\delta n)]$ decays to zero as $n \to \infty$, when $ \delta <  0.5c-1 $.
\eprf


Thus, for any given graph, a random graph does not have edit distance less than $ (0.5c-1) n$ from it. It is thus reasonable to expect for any graph reconstruction algorithm to achieve an edit distance less than $(0.5c-1) n$, since otherwise, a random choice of the output graph could achieve the same edit distance.  We now provide a lower bound on the fraction of the participating nodes such that no algorithm can reconstruct the original graph up to an edit distance less than $(0.5c-1) n$.

\begin{theorem}[Lower Bound]\label{thm:lowerbound}
For $G_n \sim \Gmsc(n,c/n)$  and any set of participants $V_n$, for any graph estimator $\hG_n$ using   (exact) shortest path distances between the participating node pairs, we have\begin{align}\nn &\Pbb[\Delta(\hG_n,G_n;V) > \delta n]\to 1, \,\,\mbox{when}\\& |V|^2< M n(0.5c- \delta -1)  \frac{\log n}{\log \log n},\end{align}  for a small enough constant $M>0$ and any $\delta <(0.5c-1)$.
\end{theorem}

Thus, no   algorithm can reconstruct $G_n$ up to edit distance $\delta n$, for  $\delta <  0.5c-1 $, if the number of participating nodes is below a certain threshold. From Lemma~\ref{lemma:rand-graphs}, almost every random graph has an edit distance greater than    $( 0.5c-1 )n$ from a given graph. Thus, when the number of participating nodes is below a certain threshold, accurate reconstruction by any algorithm  is impossible.

\noindent{\bf Remarks: }

\noindent{\em (i) }The lower bound does not require that the participating nodes are chosen uniformly and holds for any set of participating nodes of given cardinality.

\noindent{\em (ii) }The lower bound is analogous to a strong converse in information theory \cite{Cover&Thomas:book} since it says that the probability of edit distance being more a certain quantity goes to one (not just bounded away from zero).


\noindent{\em (iii) }The result is valid even for the scenario where second shortest path distances are   used since the maximum second shortest path distance  is also $O(\log n)$.

\noindent{\em (iv) }We have earlier shown that our algorithms $\estone$ and $\esttwo$ have good performance under a sub-linear number of participants. Closing the gaps in the exponents between lower bound and achievability  is of interest.

\noindent{\bf Proof Ideas: }

The proof is based on information-theoretic covering type argument, where cover the range of the estimator with random graphs of high likelihood. Using bounds on binomial distribution, we obtain the desired lower bound.

\subsection{Non-Identifiability of General Topologies}\label{sec:nonidentifiable}

\begin{figure}[t]\subfloat{\begin{minipage}{1.4in}\centering{\bp\psfrag{a}[l]{$a$}
\psfrag{a}[l]{$a$}\psfrag{e}[l]{$w$}\psfrag{h1}[l]{$h_1$}\psfrag{h2}[l]{$h_2$}
\psfrag{b}[l]{$b$}\psfrag{c}[l]{$u$}\psfrag{d}[l]{$v$}
\includegraphics[width=1.4in]
{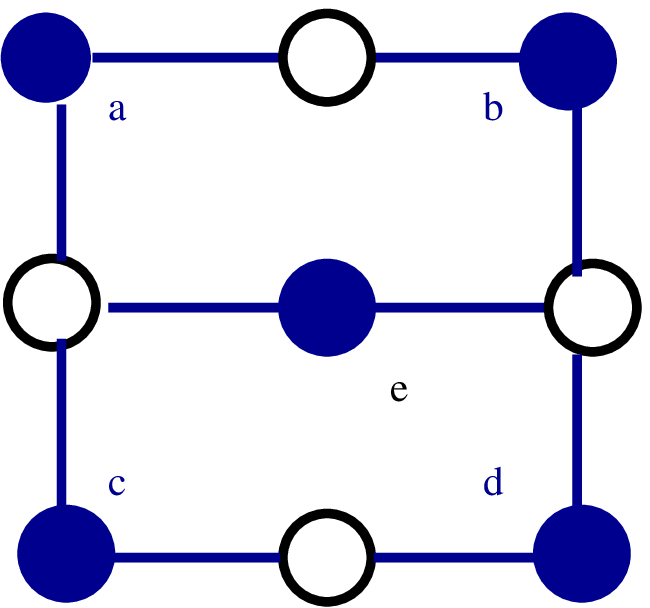}\ep}\end{minipage}}\hfil
\subfloat{\begin{minipage}{1.4in}\centering{\bp
\psfrag{a}[l]{$a$}\psfrag{e}[l]{$w$}\psfrag{h1}[l]{$h_1$}\psfrag{h2}[l]{$h_2$}
\psfrag{b}[l]{$b$}\psfrag{c}[l]{$u$}\psfrag{d}[l]{$v$}
\includegraphics[width=1.4in]
{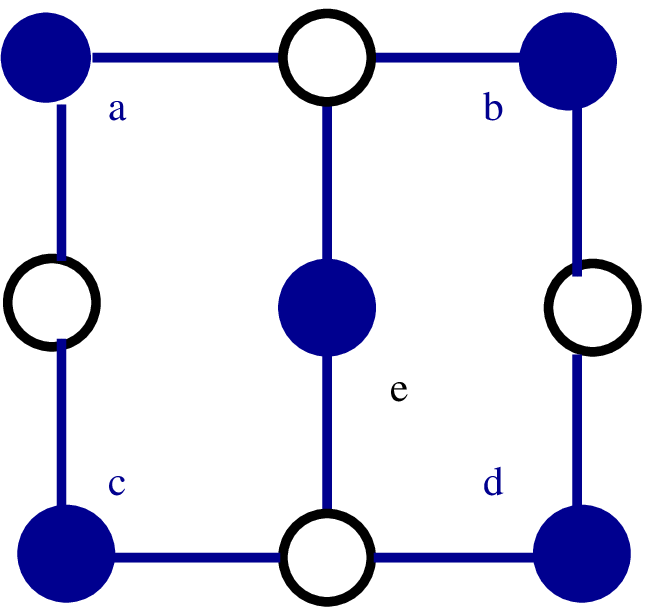}\ep}\end{minipage}}\caption{Example
of  two graphs with unit lengths where nodes $a,b,u,v,w$ are participating. Even under all path length information between the participating nodes, the two graphs cannot be distinguished. }\label{fig:nonidentifiable}\end{figure}
Our proposed algorithms  require the knowledge of shortest and second shortest path distances. Performance analysis reveals that the knowledge of second shortest path can greatly improve the accuracy of topology discovery for random graphs. We now address the question if this can be accomplished in general.

To this end, we provide a counter-example in Fig.\ref{fig:nonidentifiable}, where a significant fraction of nodes are participating, and    we are given distances along all the paths between the participants; yet, the topology cannot be correctly identified by any algorithm. This reveals a fundamental non-identifiability of general topologies using only a subset of participating nodes.

\subsection{Relationship to Phylogenetic Trees}

We note some   key differences between the phylogenetic-tree model~\cite{Durbin} and the additive delay model employed in this paper. In phylogenetic trees,  sequences of extant species are available, and the unknown phylogenetic  tree is to be inferred from these sequences. The phylogenetic-tree models the series of mutations occurring as the tree progresses and new species are formed.  Efficient algorithms with low sample complexity have been proposed for phylogenetic-tree reconstruction, e.g., in~\cite{erdos99,daskalakis06}.

In the phylogenetic-tree model, the correlations along the phylogenetic tree decay exponentially with the number of hops. This implies that   long-range correlations (between nodes which are far away) are ``hard'' to estimate, and require large number of samples (compared to the size of the tree) to find an accurate estimate. However, under the delay model, the delays are additive along the edges, and  even long-range delays can be shown to be ``easy'' to estimate. Hence, the delay model does not require the more sophisticated techniques developed for phylogenetic-tree reconstruction (e.g.,~\cite{daskalakis06}), in order to  achieve low sample complexity. However, the presence of cycles complicates the analysis for delay-based reconstruction of random graphs.   Moreover, we developed algorithms when additional information is available in the form of   second shortest-path distances. Such information cannot be obtained from  phylogenetic data. We demonstrated that this additional information leads to drastic improvement in the accuracy of random-graph discovery.

\section{Conclusion}

In this paper, we considered discovery of sparse random graph topologies using a sub-linear number of uniformly selected participants. We proposed local quartet-based algorithms which exploit the locally tree-like property of sparse random graphs. We first showed that a sub-linear edit-distance guarantee can be obtained using end-to-end measurements along the shortest paths between a sub-linear number of participants. We then considered the scenario where   additionally, second shortest-path measurements are available, and showed that consistent topology recovery is feasible using only a sub-linear number of participants. Finally, we establish a lower bound on the edit distance achieved by any algorithm for a given number of participants. Our algorithms are simple to implement, computationally efficient and have low sample complexity.

There are many interesting directions to explore.  Our algorithms require the knowledge of the bounds on the delay variances (i.e., edge lengths), and   algorithms which remove these requirements can be explored. Our algorithms are applicable for other locally tree-like graphs as well, while the actual performance indeed depends on the model employed. Exploring how the reconstruction performance changes with the  graph model is of interest. In many networks, such as peer-to-peer networks, there is a high churn rate and the nodes join and leave the networks, and it is of interest to extend our algorithms to such scenarios. Moreover, we have provided reconstruction guarantees in terms of edit distance with respect to the minimal representation, and plan to analyze reconstruction of other graph-theoretic measures such as the degree distribution, centrality measures, and so on. While we have assumed uniform sampling, other strategies (e.g., random walks) need to analyzed. We plan to implement the developed algorithms developed on real-world data.
%

\subsubsection*{Acknowledgements}
The authors thank the anonymous reviewers for comments which significantly improved this paper. The first author is supported in part by the setup funds at UCI and the AFOSR Award FA9550-10-1-0310.

\begin{appendix}

\section{Properties of Random Graphs}\label{randomgraphs}
We first note the number of cycles in random graphs.

\bl[Cycles in Erd\H{o}s-R\'{e}nyi Random Graphs]\label{lemma:randomgraphold}
In $G_n\sim \Gmsc(n, \frac{c}{n})$, the expected number of cycles of lengths $l$ is $O(c^l)$. Moreover, the number of two overlapping cycles of length $l$, denoted by $H_{l}$,    satisfies
\beq
\label{eqn:M}\Ebb[ N_{H_{l}}]= O(n^{-1}  c^{2l+1}).
\eeq\el

Thus, there are a.a.s. no overlapping cycles of length less than $\frac{(1-\delta)\log n}{2 \log c}$ for $\delta>0$.


\bprf The proof is along the lines of \cite[Cor. 4.9]{Bollobas:book}, but we specialize it for cycles. By counting argument, the expected number of cycles is given by \[ \Ebb[  N_{\Cc(l)}] = \binom{n}{l}\frac{l!}{2 l} \left(\frac{c}{n} \right)^{ l}=O(c^l).\]  Let number of vertices in $H$, be $|v(H_{l})| =s$ with $l< s\leq 2l$. Note that the number of edges $|H_{ l}|\geq s+1$ to be overlapping cycles. Hence, \begin{align}\nn
\Ebb[N_{H_{l}}] &\leq \binom{n}{s} (s!) \, \left(\frac{c}{n} \right)^{s+1}\\ &= O(n^{-1}  c^{s+1}),\nn \end{align} and we obtain the desired result.\eprf

Since we are dealing with the minimal representative $\tilG_n$ obtained by contracting nodes of degree $<3$ in the original random graph $G_{n'}$, we need to derive its  distribution. First note that
 $n = \Theta(n')$ a.a.s., where $n'$ is the number of nodes in the original graph. On lines of~\cite[Lemma 4.4]{benjamini2006mixing} and~\cite[Lemma 5.1]{benjamini2006mixing},  conditioned on $n$ nodes  in the minimal representative, the resulting graph is Erd\H{o}s-R\'{e}nyi, conditioned on the event that the minimum degree is at least three and denote this distribution as $\Gmsc'(n, \frac{c}{n})$.


We  now obtain a lower bound on the size of the neighborhood in $l$ hops in $\Gmsc'(n,c/n)$. Let $\Gamma_l(i)$ denote the set of nodes at graph distance $l$ from node $i$ in $\Gmsc'(n, \frac{c}{n})$.
 We have the following result.

\begin{lemma}[Neighborhood in $\Gmsc'(n,\frac{c}{n})$]\label{lemma:nbd}For each node $i $ in graph $\tilG_n \sim \Gmsc'(n , \frac{c}{n})$, with probability at least $1- o(1/n)$,
\beq |\Gamma_l(i)| \geq \frac{1}{(\sqrt{c}-1)^2}c^{l-l_0} \log n ,\eeq for all $l_0\leq l\leq \frac{3 \log n}{5 \log c}$, where \beq\label{eqn:l0} l_0\leq  \frac{ \log \frac{9 \log n}{(\sqrt{c}-1)^2}}{\log 3}.\eeq\end{lemma}

\bprf The proof is along the lines of~\cite[Lemma 6]{Chung&Lu:01AAM} but with modification to account for the minimum degree of three. Let $l_0$ denote the first time when \beq |\Gamma_{l_0}(i)| \geq \frac{9 \log n}{(\sqrt{c}-1)^2}.\eeq Since the minimum degree is at least three, $l_0$ is given by \eqref{eqn:l0}. The rest of the proof proceeds along the lines of~\cite[Lemma 6]{Chung&Lu:01AAM}.
  \eprf

We now provide bounds on the number   cycles in $\Gmsc'(n, \frac{c}{n})$. Let \beq \xi(c):=1- e^{-c} - c e^{-c} - 0.5 c^2 e^{-c}.\label{eqn:xi2}\eeq

\bl[Cycles in $\Gmsc'(n, \frac{c}{n})$]\label{lemma:randomgraph}
In $\tilG_n\sim \Gmsc'(n, \frac{c}{n})$, the expected number of cycles of lengths $l$ is \beq \Ebb'[N_{C_l}]= O\left(\left(\frac{c}{\xi(c)}\right)^l\right).\eeq Moreover, the number of two overlapping cycles of length $l$, denoted by $H_{l}$,    satisfies
\beq
\label{eqn:M'}\Ebb'[ N_{H_{l}}]= O\left(\frac{n^{-1}  c^{2l+1}}{\xi(c)^l}\right).
\eeq\el

\bprf Let $\Ebb'[N_{C_l}]$ denote the expected number of cycles of length $l$ in random graph $\Gmsc'(n,c/n)$ and let $\Ebb[N_{C_l}]$ denote the corresponding number in Erd\H{o}s-R\'{e}nyi random graph $\Gmsc(n,c/n)$. Let $\Lambda_n(l)$ denote the event that all given $l$   nodes have degree at least three in $\Gmsc(n,c/n)$, and let $\Phi_n(l)$ denote the event that all given $l$   nodes have degree at least three in $\Gmsc(n,c/n)$ and have edges only to nodes other than the given $l$ nodes.
Thus, we have that \[\Ebb'[N_{C_l}]=\Ebb[N_{C_l}|\Lambda(l)]=\frac{\Ebb[N_{C_l} 1_{\Lambda(l)}]}{\Pbb[\Lambda(l)]} \leq \frac{\Ebb[N_{C_l}]}{\Pbb[\Lambda(l)]} =O\left(\left(\frac{c}{\xi(c)}\right)^l\right),\] where $1$ denotes indicator event
and \begin{align} \Pbb[\Lambda_n(l)]\geq \Pbb[\Phi_n(l)]=(\Pbb[\Phi_n(1)])^l   \overset{n \to \infty}{=} \xi(c)^l,\end{align} where the last result is from the fact that the asymptotic degree distribution of a node is the Poisson distribution. Similarly we have the other result on number of overlapping cycles.
\eprf
\section{Proof of Theorem~\ref{thm:estone}}\label{proof:estone}

To prove the reconstruction guarantees for $\estone$ algorithm, we first characterize ``good'' events which lead to accurate addition of edges in each step of $\estone$ algorithm. We then bound the number of ``bad'' events which leads to an edit distance guarantee between the reconstructed graph $\hG$ by $\estone$ algorithm  (under exact distances) and the minimal representation of the original graph $\tilG$.


Recall  in Section~\ref{sec:cycles}, we introduced the concept of bad quartets, where a middle hidden node is part of a  cycle of length less than $2R'$ hops in the original graph $G_n$, where $R'=Rg/f$. Such quartets have wrong edge lengths or are not discovered.
We weaken the criterion for bad quartets as those,  where a middle hidden node is part of a (generalized)   cycle of length less than $3R'$ hops. We note that this suffices to guarantee the presence of good witnesses which leads to accurate merging of the quartet under consideration. We prove this fact below.

\begin{lemma}[Correctness of $\estone$ for good quartets]
Given a minimal representation $\tilG$ and a set of observed nodes $V$, conditioned on the event that every edge in $\tilG$ is part of a short quartet (with edge lengths less than $Rg+\tau$), each short quartet is successfully and accurately merged by $\estone$ when its middle hidden node is not part of a (generalized)   cycle of length less than $3R'$ hops.
\end{lemma}

\bprf The proof proceeds by induction on the steps of $\estone$. Initially the graph is empty and since the quartet added is good, it is correct. At any step, assume that the graph $\hG$ is accurate (i.e., either the hidden nodes and paths are not yet added, or if there are added are correct). Let $Q(ab|uv)$ be the quartet to be merged with $\hG$ and let $h_1$ and $h_2$ be its two hidden nodes. If $\TreeMerge$ procedure is called by $\estone$ algorithm in this step, it is accurate since it correctly adds the quartet $Q(ab|uv)$ to $\hG$. If $\CycleMerge$ procedure is called instead, the quartet $Q(ab|uv)$ is accurately merged if the join points between the existing paths in $\hG$ and the new paths to be created are correct. Note that the distance between hidden nodes $h_1$ and $h_2$ to be added and the join points to be inferred is at most $R'$ hops. Since each of the join points  is part of the short quartet, these short quartets are part of the witness set $\Wc$. If the witness quartets are not part of cycles of length less than $2R'$ hops, then they are guaranteed to be of the correct length and the join points for $Q(ab|uv)$ are correctly discovered. This implies that the middle nodes in $Q(ab|uv)$ are required to be not part of generalized short cycles of length less than $3R'$. Thus, the graph $\hG$ is accurate upon merging $Q(ab|uv)$. This implies the correctness of $\estone$ at each step and thus, the above statement holds. \eprf

Thus, the above result implies that the errors occur due to the following events: let $\Ec_1(e;\tilG_n , V_n)$ denote the event that the edge $e$ is not a middle edge in any short quartet. Let $\Ec_2(v;\tilG_n,V_n)$ denote the event that the node $v$ is the middle node of a bad short quartet, and let $K_v$ denote the number of such bad short quartets (with participating nodes as end points and $v$ as one of the middle nodes).
The   edit distance satisfies
\begin{align} \Delta(\hG_n,\tilG_n;V_n)  \leq\sum_{v \in \tilG_n}&  (\Deg(v;\tilG_n)+6K_v) \,\Ibb[ \Ec_2(v)] + 2n^2\sum_{e\in G_n} \Ibb[\Ec_1(e)] \label{eqn:editdistance}.\end{align}
This is because under event $\Ec_2(v)$,   $v$ is the middle node of a bad quartet, either it is not reconstructed, in which case, it contributes an edit distance of at most $\Deg(v)$, or the  bad quartet  is reconstructed with wrong edge lengths. In this case, it amounts to adding  three wrong edges and not reconstructing the three correct edges. Thus, the edit
distance is at most $6K_v$, where $K_v$ is the number of bad quartets having $v$ as a middle node under this event.
For event $\Ec_1(e)$, where there is no short quartet containing $e$ as a middle edge, we use the trivial bound on the edit distance as $2n^2$.

%


For the event $\Ec_1(e)$, we have
\[ \Pbb[\Ec_1(e; G_n,V_n)]\leq 2\max_{v\in V} \Pbb[|V_n\cap B_{R/2}(v;\tilG_n)|<2],\] since $ \Ec^c_1(e; G_n,V_n)]= \{|V_n\cap B_{R/2}(v_1;\tilG_n)\geq2\}\cap \{|V_n\cap B_{R/2}(v_2;\tilG_n)\geq2\} $, where $v_1$ and $v_2$ are the endpoints of $e$. We now have \[\Pbb\left[|V_n\cap B_{R/2}(v;\tilG_n)|<2\Big||B_{R/2}(v;\tilG_n)|\geq k\right] \leq (1-\rho)^k\left(1+\frac{\rho}{1-\rho}\right).\]
We have a lower bound on $|B_{R/2}(v;\tilG_n)|$ from Lemma~\ref{lemma:nbd}.
Hence, for \beq \label{eqn:Rbound}
R_{\min}\leq R\leq   R_{\max},\eeq where $R_{\min}$ and $R_{\max}$ are given by  \eqref{eqn:Rmax}, with probability $1-o(n^{-1})$,  we have
 \[ |B_{R/2}(v)| \geq (1-\delta)c^{(R-R_{\min})/2},\] for some constant $\delta>0$.  Thus,   \beq  \Pbb[|V\cap B_{R/2}(v)|<2]\leq  (1-\rho)^{(1-\delta) c^{(R-R_{\min})/2}}(1+\frac{\rho}{1-\rho}) .\eeq

For the second event $\Ec_2$, that the edge $v$ is a part of a bad quartet, this occurs when it is part of a (generalized) cycle of length less than $3R'$,    \[ \Pbb[\Ec_2(v; \tilG_n,V_n)]= \Pbb[v \in \Cc(3R';\tilG_n)],\] where $R' = Rg/f= \frac{\gamma g \log n}{f\log c}$.
We have from Lemma~\ref{lemma:randomgraph}, \[\Pbb[e \in \Cc(3R';\tilG_n)]
= O\left(\left(\frac{c}{\xi(c)}\right)^{3R'} \right).\]

The number of bad short quartets $K_v$ satisfies
\[ K_v = \tilde{O}(\rho^4 c^{2R'} ),\]since $K_v \leq |V_n\cap B_{R'/2}(v;\tilG_n)|^4$ and
\[ \Ebb[|V_n\cap B_{R'/2}(v;\tilG_n)|]\leq  \rho \left( \frac{c}{\xi(c)}\right)^{R'/2}, \] and using Chernoff bounds, we have $|V_n\cap B_{R'/2}(v;G_n)| = \tilde{O}(\rho (c/\xi(c))^{R'/2} )$ with probability $1-o(n^{-1})$.

Thus, the expected edit distance  is \begin{align}\nn \Ebb[\Delta(\hG_n,\tilG_n;V_n)]&= O(n^4 (1-\rho)^{(1-\delta)c^{(R-R_{\min})/2}})\nn\\& + \tilde{O}\left(\left(\frac{c}{\xi(c)}\right)^{5R'}\rho^4 \right)\label{eqn:twoterms}.\end{align}
 Let $\rho_n = n^{-\beta}$.
We have\[(1-\rho)^{\Theta(c^{(R-R_{\min})/2})}= O(\exp[-n^{ \gamma/2-\beta}]) ,\] 
when  $\gamma >2 \beta$ and $\gamma := \frac{\log c}{\log n}(R-R_{\min})$.  When  $(c/\xi(c))^{R'} = n^{\mu g/f }$, the second term in \eqref{eqn:twoterms} is $\tilde{O}(n^{5 \mu g/f - 4 \beta })$, and is the dominant error event. Thus, the expected edit distance is
\[\Ebb[\Delta(\hG_n,\tilG_n;V_n)] = \tilde{O}(n^{5 \mu g/f - 4 \beta}).\] By Markov inequality, we have the result. \qed

\section{Proof of Theorem~\ref{thm:esttwo}}\label{proof:esttwo}

The proof follows the lines of proof of Theorem~\ref{thm:estone}. It is easy to note that each step of $\esttwo$ succeeds and accurately merges a candidate quartet $Q(ab|uv)$ when the quartet has at most one (generalized) cycle of length less than $3R'$. This is because in this case, the join points of the quartet $Q(ab|uv)$ in $\hG$ can be inferred using shortest and second shortest paths. As in Theorem~\ref{thm:estone}, we require that all edges be part of short quartets.
Thus, We again have error events $\Ec_1$ and $\Ec_2$ which lead to a bound on edit distance \eqref{eqn:editdistance}. As before,  let $\Ec_1(v;\tilG_n , V_n)$ denote the event that the node $v$ is not a middle node in any short quartet and $\Ec_2(v)$ is the    event that the node $v$ is the middle node of a bad short quartet. However, now, the definition of a bad quartet is different: it occurs only when node $v$ part of at least two overlapping generalized cycles, both of length less than $3R'$, and denote such structures as $H_{3R'}$.

The analysis for $\Ec_1$ is same as in proof of Theorem~\ref{thm:estone}. For $\Ec_2$,  from Lemma~\ref{lemma:randomgraph}, we have,
\[  \Pbb[v\in H_{3R'}] = O(n^{-2}  (\frac{c}{\xi(c)})^{6R'}).\]

Thus, the expected edit distance  is \begin{align}\nn \Ebb[\Delta(\hG_n,\tilG_n;V_n)]&= O(n (1-\rho)^{ c^{(R-R_{\min})/2}})\nn\\& +O(n^{-1}(\frac{c}{\xi(c)})^{8R'}\rho^4  )\nn.\end{align}
Thus, we have the desired result.
\qed

\section{Proof of Theorem~\ref{thm:sample}}\label{proof:sample}

The proof follows the lines of sample complexity results in \cite{Bhamidi&etal:09Rand}. From \cite[Proposition 1]{Bhamidi&etal:09Rand}, we have concentration bounds for delays (distances) under $m$ samples as\[ \Pbb[|\hl^m(i,j)- l(i,j)|>\epsilon] \leq 2 \exp[- \frac{m \epsilon^2}{M R^3}],\] for any $\epsilon>0$, some constant $M>0$, and for all $i,j\in V_n$. Taking union bound over all node pairs, we see that when $m = \Omega(\poly(\log n))$, we have concentration of all the distances and we have the desired result.
\qed

\section{Proof of Theorem~\ref{thm:lowerbound}}\label{proof:lowerbound}

 We use a covering argument for obtaining the lower bound, inspired by \cite[Thm. 1]{Bresler&etal:Rand}. For reconstructed graph $\widehat{G}$ using shortest path distances between $O(|V(G)|^2/2)$ node pairs, the range $\Rc(\widehat{G})$ of the estimator is bounded by \[ |\Rc(\hG)|\leq (\Diam(G))^{|V(G)|^2/2},\] since the delay variances on the edges are assumed to be known exactly, and the shortest path can range from $1$ to $\Diam(G)$. For  $G\sim \Gmsc(n, c/n)$, the diameter\footnote{The diameter of $G(n,\frac{c}{n})$ is $C(c) \log n$ \cite{Fernholz&Ramachandran:07Rand}, where $C(c)= \frac{1}{\log c} + \frac{2}{c} + O(\frac{\log c}{c^2})$ as $c \to \infty$.} is  $O(\log n)$ w.h.p.
Let $\Sc(\hG;\delta n)$ denote all the graphs which are within edit distance of $\delta n$ of the graphs in range $\Rc(\hG)$ \[ \Sc(\hG;\delta n):= \{ F: \Delta(F,G' ) \leq \delta n, \mbox{ for some }G'\in \Rc(\hG)\}.\] Thus, using \eqref{eqn:Dc_def} and \eqref{eqn:num_graphs_edit_distance}, \[ |\Sc(\hG;\delta n)| \leq \bigcup_{G'\in \Rc(\hG)}|\Dc(G';\delta n)|\leq |\Rc(\hG)| n^{(\delta +1)n} 3^{\delta n}.\]For the original graph $G\sim \Gmsc(n,c/n)$, we have the required probability \begin{align}\nn \Pbb[\Delta(\hG,G;V) > \delta n] &= \sum_{g\in \Sc^c} \Pbb[\Delta(\hG,g;V)] > \delta n] \Pbb(G=g)\nn\\&+ \sum_{g\in \Sc } \Pbb[\Delta(\hG,g;V)] > \delta n] \Pbb(G_n=g)\nn   \\
\nn &\ge \sum_{g\in \Sc^c} \Pbb[\Delta(\hG,g;V)> \delta n]\Pbb(G_n=g)   \\
&\overset{(a)}{=} \sum_{g\in \Sc^c}  \Pbb(G_n=g)  \nn \\
&\overset{(b)}{=} 1-\sum_{g\in \Sc}  \Pbb(G_n=g),  \label{eqn:1minusbound}
\end{align}
where equality (a) is due to the fact that $\Pbb[\Delta(\hG,g;V)> \delta n]=1$ for all $g\in\Sc^c$ and  (b) is due to $\sum_{g\in\Sc} \Pbb(G=g)+\sum_{g\in\Sc^c} \Pbb(G=g)=1$. From \eqref{eqn:1minusbound}, it suffices to  provide an asymptotic upper bound for the term $\Upsilon:=\sum_{g\in \Sc}  \Pbb(G=g)$.   Furthermore, let $e_g \in \{1,\ldots,\binom{n}{2}\} $ denote the number of edges in the graph $g\in\Gc_n$. Then,
\begin{equation}
\Pbb(G=g) = \left( \frac{c}{n}\right)^{e_g} \left( 1-\frac{c}{n}\right)^{\binom{n}{2}-e_g}. \label{eqn:probg}
\end{equation}We have the general result that for graphs $g_1,g_2\in\Gc_n$
\begin{equation}
e_{g_1} \le e_{g_2} \quad\Rightarrow\quad \Pbb(G=g_1)\ge \Pbb(G=g_2).
\end{equation}
Define
\begin{equation}
z := \min \left\{ l \in \Nbb:\sum_{k=1}^l \binom{\binom{n}{2}}{k} \ge |\Sc| \right\} \label{eqn:z_res}
\end{equation}We obtain \[ z \leq O(\frac{|V|^2 \log \log n}{\log n}) + 0.5n(\delta +1)+  o(1).\]   Thus,
\begin{align}\nn
\Upsilon &:= \sum_{g\in \Sc}  \Pbb (G=g)  \\
&\le \sum_{k=0}^{z} \binom{\binom{n}{2}}{k} \left(\frac{c}{n}\right)^k  \left(1-\frac{c}{n}\right)^{\binom{n}{2}-k} \nn \\
& \overset{(a)}{\le} \exp \left[ -\frac{4}{nc} \Big(0.5n(0.5c- \delta -1) - O\big(\frac{|V|^2\log \log n}{\log n}\big)-o(1)\Big)^2 \right] \nn
\end{align}
where inequality (a) follows from the fact that $\Pr(\mathrm{Bin}(N,q) \le k )\le \exp(-\frac{2}{Nq}(Nq-k)^2)$ for $k\le Nq$   with the identifications $N = \binom{n}{2}$, $ q = c/n$ and $k=z$, and that $\binom{n}{2}\geq (n/2)^2$. Finally, we observe from (a) that if $|V|^2< M n(0.5c- \delta -1)  \frac{\log n}{\log \log n}$ for small enough $M>0$, then   $\Upsilon\to 0$ as $n\to\infty$ and we obtain the required result.\qed
\end{appendix}


 

  \end{document}